\documentclass[journal,draftclsnofoot,onecolumn,twoside,12pt]{IEEEtran}
\usepackage{fixltx2e}
\usepackage[cmex10]{amsmath}
\usepackage{amssymb}
\usepackage{algorithmic}
\usepackage{array}
\usepackage{mdwmath}
\usepackage{mdwtab}
\usepackage{cite}
\usepackage{url}
\usepackage{color}
\usepackage{xcolor}
\usepackage{gensymb}
\usepackage{textcomp}

\ifCLASSINFOpdf
  \usepackage[pdftex]{graphicx}
  \graphicspath{{./input/}}
  \DeclareGraphicsExtensions{.pdf,.png,.jpg,.jpeg}
\else
  \usepackage[dvips]{graphicx}
  \graphicspath{{./input/}}
  \DeclareGraphicsExtensions{.eps}
\fi
\usepackage{pgfplots}
\pgfplotsset{compat=newest}
\pgfplotsset{/tikz/every mark/.append style={solid}}
\usepackage[caption=false,font=footnotesize]{subfig}
\renewcommand{\baselinestretch}{1.64}
\begin{document}

\newcommand{\hblack}[1]{\textcolor{black}{#1}}
\newcommand{\hred}[1]{\textcolor{red}{#1}}
\newcommand{\hblue}[1]{\textcolor{blue}{#1}}
\newcommand{\hgreen}[1]{\textcolor{green}{#1}}

\newcommand{\mc}[3]{\multicolumn{#1}{#2}{#3}}
\newcommand{\mr}[3]{\multirow{#1}{#2}{#3}}

\newcommand{\chalab}[1]{\label{cha:#1}}
\newcommand{\seclab}[1]{\label{sec:#1}}
\newcommand{\applab}[1]{\label{app:#1}}
\newcommand{\figlab}[1]{\label{fig:#1}}
\newcommand{\tablab}[1]{\label{tab:#1}}
\newcommand{\eqnlab}[1]{\label{eqn:#1}}

\newcommand{\charef}[1]{Chapter~\ref{cha:#1}}
\newcommand{\secref}[1]{Section~\ref{sec:#1}}
\newcommand{\appref}[1]{Appendix~\ref{app:#1}}
\newcommand{\figref}[1]{Fig.~\ref{fig:#1}}
\newcommand{\tabref}[1]{Table~\ref{tab:#1}}
\newcommand{\eqnref}[1]{(\ref{eqn:#1})}

\newcommand{\sizecorr}[1]{\makebox[0cm]{\phantom{$\displaystyle #1$}}}

\newcommand{\op}[1]{\operatorname{#1}}
\newcommand{\vm}[1]{\mathbf{#1}}
\newcommand{\set}[1]{\mathcal{#1}}
\newcommand{\ie}{,\ i.e.,\ }
\newcommand{\eg}{,\ e.g.,\ }
\newcommand{\dB}{\text{dB}}
\newcommand{\F}[3]{\operatorname{F}_{#1 \rightarrow #2}\left\{ #3 \right\}}
\newcommand{\Fi}[3]{\operatorname{F}^{-1}_{#1 \rightarrow #2}\left\lbrace #3 \right\rbrace}
\newcommand{\E}[1]{\operatorname{E}\left\lbrace #1 \right\rbrace}
\newcommand{\Earg}[2]{\operatorname{E}_{#1}\left\lbrace #2 \right\rbrace}
\newcommand{\var}[1]{\operatorname{var}\left\lbrace #1 \right\rbrace}
\newcommand{\cov}[2]{\operatorname{cov}\left\lbrace #1,#2 \right\rbrace}
\newcommand{\Prob}[1]{\operatorname{P}\left( #1 \right)}
\newcommand{\minim}[2]{\min_{#1}\left\lbrace #2 \right\rbrace}
\newcommand{\maxim}[2]{\max_{#1}\left\lbrace #2 \right\rbrace}
\newcommand{\vect}[1]{\operatorname{vec}\left\lbrace #1 \right\rbrace}
\newcommand{\unvect}[1]{\operatorname{unvec}\left\lbrace #1 \right\rbrace}
\newcommand{\Ndist}[1]{\mathcal{N}\left( #1 \right)}
\newcommand{\CNdist}[1]{\mathcal{CN}\left( #1 \right)}
\newcommand{\tr}[1]{\operatorname{tr}\left\lbrace #1 \right\rbrace}
\newcommand{\diag}[1]{\operatorname{diag}\left\lbrace #1 \right\rbrace}
\newcommand{\rank}[1]{\operatorname{rank}\left\lbrace #1 \right\rbrace}
\newcommand{\re}[1]{\Re \left\lbrace #1 \right\rbrace}
\newcommand{\im}[1]{\Im \left\lbrace #1 \right\rbrace}
\newcommand{\sinc}[1]{\operatorname{sinc}\left( #1 \right)}
\newcommand{\logmod}[1]{\log \left( #1 \right)}
\newcommand{\logdet}[1]{\log \operatorname{det}\left( #1 \right)}

\newcommand{\opH}{\operatorname{H}}
\newcommand{\IBF}{I_{\text{BF}}}
\newcommand{\IBFa}{I_{\text{BF,a}}}

\definecolor{lightblue}{rgb}{0.2,0.6,1}
\definecolor{darkgreen}{rgb}{0.2,0.7,0.2}
\pgfplotscreateplotcyclelist{mylist}{
 {black},
 {lightblue},
 {red},
 {darkgreen},
}
\pgfplotscreateplotcyclelist{mylist_markers}{
 {black,mark=+},
 {lightblue,mark=o},
 {red,mark=x},
 {darkgreen,mark=diamond},
}
\pgfplotscreateplotcyclelist{mylist_approx}{
 {black},
 {black, dashed},
 {lightblue},
 {lightblue, dashed},
 {red},
 {red, dashed},
 {darkgreen},
 {darkgreen, dashed},
}

\pgfplotscreateplotcyclelist{mylist_DP_approx}{
 {red},
 {red, dashed},
 {darkgreen},
 {darkgreen, dashed},
}
% \bstctlcite{BSTcontrol}

\title{Dual-Polarized Ricean MIMO Channels: Modeling and Performance Assessment}

\author{Adrian~Ispas\IEEEmembership{,~Student~Member,~IEEE}, Xitao~Gong\IEEEmembership{,~Student~Member,~IEEE}, Christian~Schneider, Gerd~Ascheid\IEEEmembership{,~Senior~Member,~IEEE}, and~Reiner~Thom\"a\IEEEmembership{,~Fellow,~IEEE}%
\thanks{This work was supported by the Ultra high-speed Mobile Information and Communication (UMIC) research centre. Parts of this work were presented at the IEEE Global Communications Conference (GLOBECOM), Anaheim, CA, USA, December 2012. %, see \cite{Ispas_Ricean_DP}.}%
Available: http://www.ice.rwth-aachen.de/fileadmin/publications/Ispas2012GLOBECOM.pdf}%
\thanks{Adrian Ispas, Xitao Gong, and Gerd Ascheid are with the Chair for Integrated Signal Processing Systems, RWTH Aachen University, Templergraben 55, 52056 Aachen, Germany; tel.: +49-241-80-\{27873, 27885, 27882\}; fax: +49-241-80-22195; email: \{ispas, gong, ascheid\}@iss.rwth-aachen.de.}%
\thanks{Christian Schneider and Reiner Thom\"a are with the Institute for Information Technology, Ilmenau University of Technology, PSF 100 565, 98684 Ilmenau, Germany; tel.: +49-3677-69-\{2622, 1397\}; email: \{christian.schneider, reiner.thomae\}@tu-ilmenau.de.}}

% \markboth{IEEE Transactions on ...,~Vol.~xx, No.~x, xxx~xxxx}%
% {Ispas \MakeLowercase{\textit{et al.}}: ...}
% \markboth{IEEE Transactions on Communications}%
% {Submitted paper}



\maketitle

\vspace{-2cm}
\begin{abstract}
\vspace{-0.2cm}
In wireless communication systems, dual-polarized (DP) instead of single-polarized (SP) multiple-input multiple-output (MIMO) transmission is used to improve the spectral efficiency under certain conditions on the channel and the signal-to-noise ratio (SNR). In order to identify these conditions, we first propose a novel channel model for DP mobile Ricean MIMO channels for which statistical channel parameters are readily obtained from a moment-based channel decomposition. Second, we derive an approximation of the mutual information (MI), which can be expressed as a function of those statistical channel parameters. Based on this approximation, we characterize the required SNR for a DP MIMO system to outperform an SP MIMO system in terms of the MI. Finally, we apply our results to channel measurements at $2.53$~GHz. We find that, using the proposed channel decomposition and the approximation of the MI, we are able to reproduce the (practically relevant) SNR values above which DP MIMO systems outperform SP MIMO systems.
\end{abstract}

\vspace{-0.2cm}
\begin{IEEEkeywords}
\vspace{-0.2cm}
Channel models, MIMO, performance evaluation, Rician channels
\end{IEEEkeywords}

\section{Introduction}

\IEEEPARstart{M}{ultiple}-input multiple-output (MIMO) transmission is by now a well established technique to enhance the spectral efficiency over wireless channels. While commonly antennas with the same polarization are considered for MIMO systems, the use of dual-polarized (DP) antennas is known to offer advantages in terms of the spectral efficiency under certain conditions on the channel and the signal-to-noise ratio (SNR). Besides being able to improve the spectral efficiency, DP antennas allow for compact MIMO systems with co-located antennas due to the strong decorrelation over orthogonal polarizations.

In order to understand the influence of channel properties and the SNR on the spectral efficiency, channel models are commonly used. The main goal of channel models is to give a simplified yet accurate representation of the effects of the channel on the transmitted signal. They thus allow to replace the use of sophisticated channel measurements that are specific to a measurement environment, and, furthermore, they can allow for analytical evaluations. A good overview on the modeling of DP MIMO channels can be found in \cite{Coldrey_MIMO_DP_Channels, Oestges_DP_Model_and_System_Eval, Erceg_MIMO_DP_Meas_and_Model}. Experimental results regarding DP MIMO channels are presented in\eg \cite{Oestges_DP_Model_and_System_Eval, Degli-Esposti_DP_Propagation, Quitin_Polarization_Measurements, Landmann_Polarization}. Furthermore, in \cite{Tian_DP_Orthogonality}, the orthogonality of DP MIMO channels is characterized, and, in \cite{Nabar_Diversity_Ricean_Channels}, the impact of Ricean fading channels on the diversity performance is investigated analytically.

Unfortunately, an accurate and analytically tractable modeling of DP MIMO channels is a difficult task. One has to resort to several assumptions in order to obtain analytical expressions\eg for the mutual information (MI), and thus to assess the influence of the channel on the spectral efficiency. It is known that DP MIMO systems are attractive in Ricean channels \cite{Oestges_MIMO_DP_Capacity_Metrics, Oestges_DP_Model_and_System_Eval}. However, the channel and the SNR conditions for a DP MIMO system to outperform a single-polarized (SP) MIMO system in terms of the spectral efficiency are not fully characterized and they are time-dependent. Expressions relating the statistical channel parameters to the spectral efficiency are usually limited to restrictive channel models with separable correlation\ie a Kronecker structure, and/or without a Ricean component; moreover, they often rely on asymptotic settings. For recent contributions regarding analytical expressions of the MI for Ricean channels in asymptotic settings, see \cite{Taricco_Asymptotic_Rician_KD_MI} and references therein. The dependence of the spectral efficiency of SP and DP MIMO channels on the SNR and the $K$-factor is demonstrated\eg in \cite{Coldrey_MIMO_DP_Channels} with simulated channels. The spectral efficiency of measured SP and DP MIMO channels with (instantaneous) channel state information (CSI) at the receiver (RX) only has been compared\eg in \cite{Anreddy_MIMO_DP_Capacity} with indoor measurements at $2.4$~GHz, in \cite{Kyritsi_MIMO_DP_Capacity} with indoor measurements at $1.95$~GHz, or in \cite{Erceg_MIMO_DP_vs_SP} with outdoor measurements at $2.5$~GHz. While \cite{Kyritsi_MIMO_DP_Capacity, Erceg_MIMO_DP_vs_SP} conclude that DP MIMO systems are favorable, \cite{Anreddy_MIMO_DP_Capacity} concludes that, especially for low $K$-factors, SP MIMO systems are recommended to reach higher spectral efficiencies. Therefore, as highlighted in \cite{Degli-Esposti_DP_Propagation}, it is not straightforward to decide when to use a DP instead of an SP MIMO system. We also note that SP MIMO systems would highly benefit from the availability of CSI at the transmitter (TX).

Consequently, we first aim at establishing a general channel model for SP and DP MIMO systems which is reasonably accurate, yet analytically tractable. Second, we aim at identifying the conditions on the channel and the SNR under which it is beneficial, in terms of spectral efficiency, to make use of the polarization domain for a limited number of antennas at both link ends. The reason to limit the number of simultaneously used antennas is that it is desirable to keep a low number of radio frequency chains since they are expensive components in a wireless system. One can then perform antenna switching between differently polarized antennas\ie between SP and DP MIMO systems.

\textit{Contributions:} We detail a general modeling approach for SP and DP MIMO channels. Furthermore, we evaluate the achievable rate over such channels for the case that the TX has only statistical CSI, while the RX has instantaneous CSI. In particular, we contribute the following:
\begin{itemize}
 \item We propose a general model for SP and DP mobile Ricean MIMO channels. Furthermore, we derive a moment-based channel decomposition yielding the statistical channel model parameters from measured data.
 \item We give an approximation of the achievable rate\ie the MI, which is an explicit function of the statistical parameters of the proposed channel model. We can thus assess the influence of the statistical channel parameters on the achievable rate.
 \item We use the approximate MI to characterize the required SNR for a DP setup to outperform an SP setup. Specifically, we give a closed-form expression of such an SNR threshold for the practically relevant case of a dual-stream DP setup vs. a single-stream SP setup.
 \item We evaluate the channel decomposition and the MI for $4\times4$ SP and DP MIMO systems based on urban macrocell measurements at $2.53$~GHz. We find that the DP setup is advantageous in terms of the MI for medium- to high-$K$-factor links above a certain SNR. With the approximate evaluation of the MI, we can reproduce the crossing points between the MI of the SP and DP MIMO systems.
\end{itemize}

\vspace{-0.05cm}
\textit{Structure:} We first introduce the MIMO system model in \secref{system_model}. Then, in \secref{channel_modeling_and_decomposition}, we develop the channel model and its corresponding decomposition technique for SP and DP channels. \secref{performance_assessment} deals with the performance assessment for SP and DP MIMO transmission. In \secref{channel_measurements}, the channel measurements and the data selection are presented, before proceeding with the results in \secref{results}. Finally, we draw the conclusion in \secref{conclusion}.

\textit{Notation:} We use lowercase and uppercase boldface letters to designate vectors and matrices, respectively. For a matrix $\vm{A}$, the (element-wise) complex conjugate, the transpose, and the conjugate transpose are denoted by $\vm{A}^{\ast}$, $\vm{A}^T$, and $\vm{A}^H$, respectively. The unique Hermitian positive semidefinite square root of a Hermitian positive semidefinite matrix $\vm{A}$ is represented by $\vm{A}^{\frac{1}{2}}$. For the matrix $\vm{A}$, $\tr{\vm{A}}$, $\rank{\vm{A}}$, and $\lambda_{\text{max}}(\vm{A})$ denote the trace, the rank, and the maximal eigenvalue, respectively. For two matrices $\vm{A}$ and $\vm{B}$, $\vm{A} \odot \vm{B}$ is the Hadamard (element-wise) product and $\vm{A} \otimes \vm{B}$ is the Kronecker product. The vectorization\ie the column-wise stacking, of the matrix $\vm{A}$ is denoted by $\vect{\vm{A}}$. The $N \times N$ identity matrix is represented by $\vm{I}_N$ and the all-zero matrix of size $N_1 \times N_2$ is denoted by $\vm{0}_{N_1,N_2}$. The real-valued $M N \times M N$ commutation matrix $\vm{K}_{M,N}$ satisfies $\vm{K}_{M,N} \vect{\vm{A}} = \vect{\vm{A}^T}$ for an $M \times N$ matrix $\vm{A}$. Consider an $M \times N$ matrix $\vm{A}$ with $k=1,\ldots,M$ and $l=1,\ldots,N$; we use $[\vm{A}]_{k,l}$ to denote the element in the $k$th row and the $l$th column of $\vm{A}$, and we define $\vm{A}^+$ such that $[\vm{A}^+]_{k,l} = \maxim{}{[\vm{A}]_{k,l},0}$ holds. Expectation is denoted by $\E{\cdot}$, $\log(\cdot)$ is the logarithm to the base $2$, and $\ln(\cdot)$ is the natural logarithm. The imaginary unit is represented by $j$.

\section{System Model}
\seclab{system_model}

We consider a MIMO channel which is characterized by time-varying and frequency-flat fading. The input-output relation for transmission from $N_{\text{TX}}$ antennas at the TX to $N_{\text{RX}}$ antennas at the RX is given at time slots $m \in \mathbb{Z}$ by the received length-$N_{\text{RX}}$ column vector
\begin{IEEEeqnarray}{rCl}
 \vm{y}[m] &=& \vm{H}[m] \vm{x}[m] + \vm{n}[m] .
 \eqnlab{received_signal}
\end{IEEEeqnarray}
The random channel matrices $\{\vm{H}[m]\}$, each of size $N_{\text{RX}} \times N_{\text{TX}}$, are jointly proper. The length-$N_{\text{TX}}$ column vectors $\{\vm{x}[m]\}$ denote the zero-mean jointly proper Gaussian transmitted vectors that are uncorrelated in time with spatial covariance matrix $\E{\vm{x}[m] \vm{x}^H[m]} = P_x \vm{Q}[m]$, $P_x>0$, and $\tr{\vm{Q}[m]} = 1$. The length-$N_{\text{RX}}$ column vectors $\{\vm{n}[m]\}$ are the white jointly proper Gaussian noise vectors in time with spatial covariance matrix $\E{\vm{n}[m] \vm{n}^H[m]} = \sigma_n^2 \vm{I}_{N_{\text{RX}}}$ and $\sigma_n^2 > 0$. The random processes $\{\vm{H}[m]\}$, $\{\vm{n}[m]\}$, and $\{\vm{x}[m]\}$ are assumed to be mutually independent. For ease of exposition, we define the (nominal) SNR $\rho = P_x / \sigma_n^2$. We assume the RX to have instantaneous CSI\ie the RX has knowledge of the current channel realization $\vm{H}[m]$. The TX, on the other hand, only has statistical CSI of the channel.

\section{Channel Modeling and Decomposition}
\seclab{channel_modeling_and_decomposition}

A channel model has to be accurate yet simple enough to offer insight on the influence of the relevant channel parameters on the system performance. Several approaches to model the channel exist; they can be mainly classified in physical and analytical models \cite{Almers_Survey_Channels}. We choose the popular correlation-based analytical modeling approach for MIMO channels which is easier to use for analytical evaluations and which requires statistical parameters that are, in general, readily available from measurement data. Correlation-based analytical models can contain a term representing line-of-sight (LOS) or a strong scatterer \cite{Wyne_MIMO_Measurements_Journal} for each MIMO sub-link. The amplitude of the sub-links then changes from a Rayleigh to a Ricean distributed random variable. The ratio between the power of the dominant component and the power of the remaining weaker component is referred to as the $K$-factor.

\subsection{Channel Model}
\seclab{channel_model}

It is common to represent the dominant components of the MIMO channel by a deterministic rank-one matrix \cite{Jin_MIMO_Rank1_Ricean_Capacity, Farrokhi_MIMO_Modeling_and_Processing}. While this is usually applicable for an SP MIMO system in an LOS scenario where the TX and the RX are fixed, it is not appropriate in general. This is especially true for DP MIMO systems where independent propagation along orthogonal polarizations might occur. Moreover, in the presence of a mobile terminal (MT), the dominant channel component\ie a strong scatterer or LOS, has a varying phase and as a consequence the mean of the channel is zero \cite{Wyne_MIMO_Measurements_Conf}.\footnote{Another reason for a zero-mean channel can be the consideration of channel samples at other frequencies as different channel realizations.} We thus introduce the following model for SP and DP mobile MIMO channels:
% \begin{IEEEeqnarray}{rCl}
%  \vm{H}[m] &=& \underbrace{\begin{bmatrix}
%  \bar{\vm{H}}_{\text{VV}}[m] & \bar{\vm{H}}_{\text{HV}}[m]\\
%  \bar{\vm{H}}_{\text{VH}}[m] & \bar{\vm{H}}_{\text{HH}}[m]
%  \end{bmatrix}}_{=\bar{\vm{H}}[m]} +~ \tilde{\vm{H}}[m]
%  \eqnlab{channel_model}
% \end{IEEEeqnarray}
\begin{IEEEeqnarray}{rCl}
 \vm{H}[m] &=& \underbrace{\begin{bmatrix} \bar{\vm{H}}_{\text{VV}}[m] & \bar{\vm{H}}_{\text{HV}}[m] \\ \bar{\vm{H}}_{\text{VH}}[m] & \bar{\vm{H}}_{\text{HH}}[m] \end{bmatrix}}_{=\bar{\vm{H}}[m]} + \underbrace{\begin{bmatrix} \tilde{\vm{H}}_{\text{VV}}[m] & \tilde{\vm{H}}_{\text{HV}}[m] \\ \tilde{\vm{H}}_{\text{VH}}[m] & \tilde{\vm{H}}_{\text{HH}}[m] \end{bmatrix}}_{=\tilde{\vm{H}}[m]}
 \eqnlab{channel_model}
\end{IEEEeqnarray}
where $\bar{\vm{H}}[m]$ contains the dominant contributions, which are due to LOS or strong scatterers, and $\tilde{\vm{H}}[m]$ contains the remaining contributions of the channel. The $N_{\text{TX},a} \times N_{\text{RX},b}$ sub-matrices
\begin{IEEEeqnarray}{rCl}
 \bar{\vm{H}}_{ab}[m] &=& \vm{V}_{ab}[m] \odot \vm{\Phi}_{ab}[m]
 \eqnlab{channel_model_dominant}
\end{IEEEeqnarray}
and $\tilde{\vm{H}}_{ab}[m]$ contain the sub-links with polarization $a$ at the TX and $b$ at the RX for $a,b \in \{ \text{V},\text{H} \}$. Here, V and H denote vertical and horizontal polarizations, respectively.\footnote{We note that other polarization choices\eg corresponding to a slanted scheme, are possible as well; however, we choose vertical and horizontal polarizations as they often have different propagation characteristics, see \cite{Kyritsi_MIMO_DP_Capacity} for an example in an indoor scenario.} The number of vertical-polarized (VP) and the number of horizontal-polarized (HP) antennas at the TX are given by $N_{\text{TX},\text{V}}$ and $N_{\text{TX},\text{H}}$, respectively. We thus have $N_{\text{TX},\text{V}} + N_{\text{TX},\text{H}} = N_{\text{TX}}$. The relations at the RX side are obtained analogously. In the SP case, we either use only VP or only HP antennas. In the DP case, we assume that, at both the TX and the RX, one half of the antennas is VP while the other half is HP. We split the dominant contributions into the deterministic amplitude matrix $\vm{V}_{ab}[m]$ and the random phase matrix $\vm{\Phi}_{ab}[m]$ with $[\vm{\Phi}_{ab}[m]]_{k,l} = e^{j \phi_{ab,(l-1)N_{\text{RX}}+k}[m]}$ for $k = 1, \ldots, N_{\text{RX},b}$ and $l = 1, \ldots, N_{\text{TX},a}$. The remaining weaker scatterers are represented by the zero-mean proper Gaussian matrix $\tilde{\vm{H}}[m]$\ie $\tilde{\vm{H}}_{ab}[m]$ for $a,b \in \{ \text{V},\text{H} \}$. As highlighted in \cite{Erceg_MIMO_DP_Meas_and_Model}, the challenging part is the modeling of the dependence between the phases of the dominant components $\phi_{ab,p}[m]$ for $p = 1, \ldots, N_{\text{TX},a} N_{\text{RX},b}$. We first consider all MIMO sub-links with polarization $a$ at the TX and $b$ at the RX. For $p,q = 1, \ldots, N_{\text{TX},a} N_{\text{RX},b}$, we assume
\begin{enumerate}
 \item $\phi_{ab,p}[m]$ is independent of $\tilde{\vm{H}}[m]$,
 \item $\phi_{ab,p}[m]$ is uniformly distributed over $[-\pi,\pi)$,
 \item $\varDelta_{\phi,ab}^{p,q}[m] = \phi_{ab,p}[m] - \phi_{ab,q}[m]$ is deterministic.
\end{enumerate}
The first two assumptions are commonly used, see\eg \cite{Erceg_MIMO_DP_Meas_and_Model}. However, a note is in order regarding the last assumption. As mentioned above, the contributions from the dominant components are not deterministic\eg due to the mobility of the MT. For the case that all MIMO sub-links of the same polarization combination $a$ and $b$ observe the same dominant component and that the distances between the TX, the RX, and a possible dominant scatterer are considerably larger than the array sizes, the resulting phase changes are equal for all of these sub-links. Therefore, $\varDelta_{\phi,ab}^{p,q}[m]$ is modeled as constant inside a region of constant statistical channel parameters\ie $\varDelta_{\phi,ab}^{p,q}[m]$ is deterministic. Clearly, assumption 3) is not satisfied for all antenna setups, e.g., it would not necessarily hold for a MIMO system made of directional antennas with different orientations. Therefore, for each polarization, we require the (directional) antennas at the TX and the RX to be oriented in the same direction. Using assumption 3), we can rewrite \eqnref{channel_model_dominant} as
% \begin{IEEEeqnarray}{rCl}
%  \!\!\bar{\vm{H}}_{ab}[m] &=& \vm{V}_{ab}[m] \odot \vm{\Delta}_{\vm{\phi},ab}[m] ~e^{j \phi_{ab}[m]},~a,b \in \{ \text{V},\text{H} \}
%  \eqnlab{channel_model_dominant_rewrite}
% \end{IEEEeqnarray}
\begin{IEEEeqnarray}{rCl}
 \bar{\vm{H}}_{ab}[m] &=& \vm{V}_{ab}[m] \odot \vm{\Delta}_{\vm{\phi},ab}[m] ~e^{j \phi_{ab}[m]},~a,b \in \{ \text{V},\text{H} \}
 \eqnlab{channel_model_dominant_rewrite}
\end{IEEEeqnarray}
where we defined $\phi_{ab}[m] = \phi_{ab,1}[m]$ and the deterministic matrix $\vm{\Delta}_{\vm{\phi},ab}[m] = \vm{\Phi}_{ab}[m] ~e^{-j \phi_{ab}[m]}$.

\subsection{Channel Correlation}
\seclab{channel_correlation}

Subsequently, we define full and transmit correlation matrices of the channel. Furthermore, we characterize the structure of the correlation matrices of the dominant components of the channel. The results will be needed for the channel decomposition in \secref{cm_decomposition} and the performance assessment in \secref{performance_assessment}.

\subsubsection{Full Channel Correlation Matrices}
\seclab{full_channel_correlation}

We first define the length-$N_{\text{TX}} N_{\text{RX}}$ column vectors $\vm{h}[m] = \vect{\vm{H}[m]}$, $\bar{\vm{h}}[m] = \vect{\bar{\vm{H}}[m]}$, and $\tilde{\vm{h}}[m] = \op{vec}\{ \tilde{\vm{H}}[m] \}$. The corresponding $N_{\text{TX}} N_{\text{RX}} \times N_{\text{TX}} N_{\text{RX}}$ full correlation matrices of the channel are then obtained as
\begin{IEEEeqnarray}{rCl}
 \vm{R}[m] &=& \op{E} \big\{ \vm{h}[m] \vm{h}^H[m] \big\};~
 \bar{\vm{R}}[m] = \op{E} \big\{ \bar{\vm{h}}[m] \bar{\vm{h}}^H[m] \big\};~
 \tilde{\vm{R}}[m] = \op{E} \big\{ \tilde{\vm{h}}[m] \tilde{\vm{h}}^H[m] \big\}
 \eqnlab{corr_matrix}
\end{IEEEeqnarray}
% \begin{IEEEeqnarray}{rCl}
%  \vm{R}[m] &=& \E{ \vm{h}[m] \vm{h}^H[m] }\\
%  \bar{\vm{R}}[m] &=& \E{ \bar{\vm{h}}[m] \bar{\vm{h}}^H[m] }\\
%  \tilde{\vm{R}}[m] &=& \E{ \tilde{\vm{h}}[m] \tilde{\vm{h}}^H[m] } .
%  \eqnlab{corr_matrix}
% \end{IEEEeqnarray}
respectively. Using assumption 1) in \secref{channel_model}, it immediately follows that
\begin{IEEEeqnarray}{rCl}
 \vm{R}[m] &=& \bar{\vm{R}}[m] + \tilde{\vm{R}}[m]
 \eqnlab{corr_matrix_relation}
\end{IEEEeqnarray}
holds. We can categorize the MIMO sub-links into co-polarized sub-links\ie links with VP to VP or HP to HP transmission, and into cross-polarized sub-links\ie links with VP to HP or HP to VP transmission. Depending on whether the four polarizations combinations share a dominant component or not, the rank of $\bar{\vm{R}}[m]$ can vary. We show in \appref{R_bar_rank} that generally we have $\rank{\bar{\vm{R}}[m]} \leq 4$. Since the cross-polarized sub-links are hardly affected by\eg the occurrence of LOS, we consider the practically relevant setting that only the co-polarized sub-links can be affected by dominant components. Then, it can be similarly shown that $\rank{\bar{\vm{R}}[m]} \leq 2$ has to be satisfied. Further specializing this setting to the case that the VP to VP and the HP to HP sub-links are affected by distinct dominant components with independent phase terms, it follows that $\rank{\bar{\vm{R}}[m]} = 2$ is satisfied. When all polarization combinations share a common dominant component, we have $\rank{\bar{\vm{R}}[m]} = 1$. For an SP setup, $\rank{\bar{\vm{R}}[m]} \leq 1$ holds.

\subsubsection{Transmit Channel Correlation Matrices}
\seclab{tx_channel_correlation}

The $N_{\text{TX}} \times N_{\text{TX}}$ TX correlation matrices are
\begin{IEEEeqnarray}{rCl}
 \vm{R}_{\text{TX}}[m] &=& \op{E} \big\{ \vm{H}^T[m] \vm{H}^{\ast}[m] \big\};~
 \bar{\vm{R}}_{\text{TX}}[m] = \op{E} \big\{ \bar{\vm{H}}^T[m] \bar{\vm{H}}^{\ast}[m] \big\};~
 \tilde{\vm{R}}_{\text{TX}}[m] = \op{E} \big\{ \tilde{\vm{H}}^T[m] \tilde{\vm{H}}^{\ast}[m] \big\} .\IEEEnonumber\vspace{-0.2cm}\\
 \vspace{-0.2cm}
 \eqnlab{TX_corr_matrix}
\end{IEEEeqnarray}
% \begin{IEEEeqnarray}{rCl}
%  \vm{R}_{\text{TX}}[m] &=& \E{ \vm{H}^T[m] \vm{H}^{\ast}[m] }\\
%  \bar{\vm{R}}_{\text{TX}}[m] &=& \E{ \bar{\vm{H}}^T[m] \bar{\vm{H}}^{\ast}[m] }\\
%  \tilde{\vm{R}}_{\text{TX}}[m] &=& \E{ \tilde{\vm{H}}^T[m] \tilde{\vm{H}}^{\ast}[m] } .
%  \eqnlab{TX_corr_matrix}
% \end{IEEEeqnarray}
With assumption 1) in \secref{channel_model}, we have
\begin{IEEEeqnarray}{rCl}
 \vm{R}_{\text{TX}}[m] &=& \bar{\vm{R}}_{\text{TX}}[m] + \tilde{\vm{R}}_{\text{TX}}[m] .
 \eqnlab{TX_corr_matrix_relation}
\end{IEEEeqnarray}
We are interested in the structure, or more specifically the rank, of $\bar{\vm{R}}_{\text{TX}}[m]$. To that end, we assume that $\vm{V}_{ab}[m] = \vm{v}_{\text{RX},ab}[m] \vm{v}_{\text{TX},ab}^T[m]$ and $\vm{\Delta}_{\vm{\phi},ab}[m] = \vm{d}_{\text{RX},ab}[m] \vm{d}_{\text{TX},ab}^T[m]$ with the deterministic length-$N_{\text{RX}}$ column vectors $\vm{v}_{\text{RX},ab}[m]$ and $\vm{d}_{\text{RX},ab}[m]$, and the deterministic length-$N_{\text{TX}}$ column vectors $\vm{v}_{\text{TX},ab}[m]$ and $\vm{d}_{\text{TX},ab}[m]$ hold.\footnote{Note that this decomposition only imposes a rank-one condition for each polarization combination, which is realistic when the distances between the TX, the RX, and possible dominant scatterers are large.} In \appref{R_tx_bar_rank}, we show that generally $\rank{\bar{\vm{R}}_{\text{TX}}[m]} \leq 4$ holds. In the case that only the co-polarized sub-links are affected by dominant components, we obtain $\rank{\bar{\vm{R}}_{\text{TX}}[m]} = 2$. Finally, for an SP setup, we have $\rank{\bar{\vm{R}}_{\text{TX}}[m]} = 1$.

\subsection{Channel Decomposition}
\seclab{cm_decomposition}

We now describe a simple method to separate the contributions of the dominant channel components and the remaining weaker scatterers from the channel correlation matrix. We thus aim at splitting $\vm{R}[m]$ into $\bar{\vm{R}}[m]$ and $\tilde{\vm{R}}[m]$. We note that in the mobile setting we cannot use the mean of the channel to decompose the channel into the dominant and the remaining channel components. We thus introduce a method to decompose the channel that is simple compared to high resolution parameter estimation techniques \cite{Landmann_Rimax}. The method is inspired by the well-known $K$-factor estimation in \cite{Greenstein_K-Factor_Letter}. It is suitable for both SP and DP MIMO channels.

We use the second- and fourth-order moments of the channel $\vm{R}[m] = \E{ \vm{h}[m] \vm{h}^H[m] }$ and $\vm{T}[m] = \E{ (\vm{h}[m] \vm{h}^H[m])^2 }$, respectively, to obtain a simple solution to the channel decomposition of $\vm{R}[m] = \bar{\vm{R}}[m] + \tilde{\vm{R}}[m]$ into $\bar{\vm{R}}[m]$ and $\tilde{\vm{R}}[m]$. From \appref{4th_matrix_moment}, we have the relation $\vm{T}[m] = \vm{R}[m] \tr{\vm{R}[m]} + \vm{R}^2[m] - \bar{\vm{R}}^2[m]$
% \begin{IEEEeqnarray}{rCl}
%  \vm{T}[m] &=& \vm{R}[m] \tr{\vm{R}[m]} + \vm{R}^2[m] - \bar{\vm{R}}^2[m]
%  \eqnlab{4th_matrix_moment_T}
% \end{IEEEeqnarray}
which can be reformulated as
\begin{IEEEeqnarray}{rCl}
 \bar{\vm{R}}^2[m] &=& \vm{R}[m] \tr{\vm{R}[m]} + \vm{R}^2[m] - \vm{T}[m] .
 \eqnlab{R_bar}
\end{IEEEeqnarray}
% With the eigendecomposition $\bar{\vm{R}}[m] = \bar{\vm{U}}[m] \bar{\vm{\Lambda}}[m] \bar{\vm{U}}^H[m]$, we have $\bar{\vm{R}}^2[m] = \bar{\vm{U}}[m] \bar{\vm{\Lambda}}^2[m] \bar{\vm{U}}^H[m]$. Thus, we can directly obtain the unitary eigenvector matrix $\bar{\vm{U}}[m]$ and the diagonal eigenvalue matrix $\bar{\vm{\Lambda}}[m]$ of $\bar{\vm{R}}[m]$.
With the eigendecomposition $\bar{\vm{R}}[m] = \bar{\vm{U}}[m] \bar{\vm{\Lambda}}[m] \bar{\vm{U}}^H[m]$, we can thus directly obtain the unitary eigenvector matrix $\bar{\vm{U}}[m]$ and the diagonal eigenvalue matrix $\bar{\vm{\Lambda}}[m]$ of $\bar{\vm{R}}[m]$.

\subsubsection{Dual-Polarized Channel}
\seclab{cm_decomposition_DP}

According to \secref{full_channel_correlation}, at most four eigenvalues of $\bar{\vm{R}}[m]$ are non-zero; however, only two can be highly significant and smaller eigenvalues tend to be estimated less accurately. We thus have to exercise care in choosing the number of considered eigenvalues $N_{\text{DP}}$. Subsequently, we first find an estimate of $\bar{\vm{R}}[m]$ denoted as $\check{\vm{R}}[m]$ according to \eqnref{R_bar}. We then extract the $N_{\text{DP}}$ largest eigenvalues of $\check{\vm{R}}[m]$; this step is akin to taking the best rank-$N_{\text{DP}}$ approximation of $\check{\vm{R}}[m]$ in terms of the matrix $2$-norm \cite[Th.~2.5.3]{Golub_Matrix_Computations}. Clearly, we have $N_{\text{DP}} \leq 4$. The final estimate of $\bar{\vm{R}}[m]$ is
\begin{IEEEeqnarray}{rCl}
 \bar{\vm{R}}^{(e)}[m] &=& \sum_{k=1}^{N_{\text{DP}}} c_k[m] \check{\vm{u}}_k[m] \check{\vm{u}}_k^H[m]
 \eqnlab{R_bar_estimate_DP}
\end{IEEEeqnarray}
where the vector $\check{\vm{u}}_k[m]$ denotes the eigenvector corresponding to the $k$th largest eigenvalue $\check{\lambda}_k[m]$ of $\check{\vm{R}}[m]$ for $k=1,\ldots,N_{\text{DP}}$. We now define the (positive semidefinite) estimates of $\vm{R}[m]$ and $\tilde{\vm{R}}[m]$ as $\vm{R}^{(e)}[m]$ and $\tilde{\vm{R}}^{(e)}[m]$, respectively. Moreover, we define $\breve{\vm{R}}_l[m] = \vm{R}^{(e)}[m] - \sum_{k=1}^{l} c_k[m] \check{\vm{u}}_k[m] \check{\vm{u}}_k^H[m]$ for $l=0,\ldots,N_{\text{DP}}$. The parameters $c_k[m]$ for $k=1,\ldots,N_{\text{DP}}$ are chosen such that $\tilde{\vm{R}}^{(e)}[m] = \vm{R}^{(e)}[m] - \bar{\vm{R}}^{(e)}[m]$ is positive semidefinite, see \appref{psd_condition}:
% \begin{IEEEeqnarray}{rCl}
%  c_k[m] &=& \begin{cases}
%  0,~\text{for singular}~\breve{\vm{R}}_{k-1}[m]\\
%  \minim{}{ \check{\lambda}_k^+[m], \left( \check{\vm{u}}_k^H[m] \breve{\vm{R}}_{k-1}^{-1}[m] \check{\vm{u}}_k[m] \right)^{\!-1} \! } \!\!, \text{else.}\!\!
%  \end{cases}
%  \eqnlab{c_condition_DP}
% \end{IEEEeqnarray}
\begin{IEEEeqnarray}{rCl}
 c_k[m] &=& \begin{cases}
 0 ,~ \text{for singular}~\breve{\vm{R}}_{k-1}[m]\\
 \minim{}{ \check{\lambda}_k^+[m], \left( \check{\vm{u}}_k^H[m] \breve{\vm{R}}_{k-1}^{-1}[m] \check{\vm{u}}_k[m] \right)^{-1} } , ~\text{else.}
 \end{cases}
 \eqnlab{c_condition_DP}
\end{IEEEeqnarray}
Note that some power of the dominant components corresponding to $\check{\vm{u}}_k[m]$ is transferred from $\bar{\vm{R}}^{(e)}[m]$ to $\tilde{\vm{R}}^{(e)}[m]$ whenever $c_k[m] < \check{\lambda}_k[m]$. This might occur when the estimates of the moments $\vm{R}[m]$ and $\vm{T}[m]$ are inaccurate.

\subsubsection{Single-Polarized Channel}
\seclab{cm_decomposition_SP}

From \secref{channel_model}, we know that $\bar{\vm{R}}[m]$ can at most have rank one. We thus obtain the following estimate of $\bar{\vm{R}}[m]$:
\begin{IEEEeqnarray}{rCl}
 \bar{\vm{R}}^{(e)}[m] &=& c_1[m] \check{\vm{u}}_1[m] \check{\vm{u}}_1^H[m] .
 \eqnlab{R_bar_estimate_SP}
\end{IEEEeqnarray}
The constant $c_1[m]$ is chosen as in \eqnref{c_condition_DP} to ensure the positive semidefiniteness of $\tilde{\vm{R}}^{(e)}[m]$. We can generate SP channel realizations $\vm{H}^{(g)}[m]$ based on the statistical channel parameters according to\looseness=-1
\begin{equation}
 \vect{\vm{H}^{(g)}[m]} = \sqrt{c_1[m]}~ \check{\vm{u}}_1[m] e^{j \phi} + \left( \tilde{\vm{R}}^{(e)}[m] \right)^{\frac{1}{2}} \vm{g}
 \eqnlab{channel_realizations}
\end{equation}
where $\phi$ is uniformly distributed over $[-\pi,\pi)$, and $\vm{g}$ is a zero-mean proper Gaussian random column vector of length $N_{\text{TX}} N_{\text{RX}}$ with covariance matrix $\vm{I}_{N_{\text{TX}} N_{\text{RX}}}$; $\phi$ and $\vm{g}$ are mutually independent.\looseness=-1

\section{Performance Assessment}
\seclab{performance_assessment}

With respect to the system model in \secref{system_model}, the MI between the input $\vm{x}[m]$ and the output $\vm{y}[m]$ combined with instantaneous CSI at the receiver is given in bit/channel use (bit/c.u.) by
% \begin{IEEEeqnarray}{rCl}
%  \IEEEeqnarraymulticol{3}{l}{  I \left( \vm{x}[m]; \vm{y}[m], \vm{H}[m] \right) }\IEEEnonumber\\
%  &=& \E{\logdet{\vm{I}_{N_{\text{RX}}} + \rho \vm{H}[m] \vm{Q}[m] \vm{H}^H[m]}}\IEEEnonumber\\
%  &\stackrel{(\text{a})}{=}& \E{\logdet{\vm{I}_{N_{\text{TX}}} + \rho \vm{H}^H[m] \vm{H}[m] \vm{Q}[m]}}
%  \eqnlab{MI}
% \end{IEEEeqnarray}
\begin{IEEEeqnarray}{rCl}
 I \left( \vm{x}[m]; \vm{y}[m], \vm{H}[m] \right) &=& \E{\logdet{\vm{I}_{N_{\text{RX}}} + \rho \vm{H}[m] \vm{Q}[m] \vm{H}^H[m]}}\IEEEnonumber\\
 &\stackrel{(\text{a})}{=}& \E{\logdet{\vm{I}_{N_{\text{TX}}} + \rho \vm{H}^H[m] \vm{H}[m] \vm{Q}[m]}}
 \eqnlab{MI}
\end{IEEEeqnarray}
where, in (a), we used \cite[Th.~1.3.20]{Horn_Matrix_Analysis}. Note that the MI in \eqnref{MI} is time-dependent as the channel is in general non-stationary; therefore, in a strict sense, \eqnref{MI} is not an achievable rate. Nevertheless, we use the MI \eqnref{MI} as performance measure since it has an interpretation in terms of an achievable rate in bit/channel use (bit/c.u.) for non-stationary slow- and fast-fading wireless channels \cite{Ispas_MI_ICI, Lozano_Ergodic_Modeling}.
%  We note that in the SP case the MI is invariant to the distribution of the random scalar phase of the dominant component $\phi_{aa}, a \in \{\text{V},\text{H}\}$ in \eqnref{channel_model_dominant_rewrite}, see also \eqnref{channel_realizations}.

With \appref{MI_approx}, we can state the following second-order approximation of \eqnref{MI}:
\begin{IEEEeqnarray}{rCl}
 \IEEEeqnarraymulticol{3}{l}{ I \left( \vm{x}[m]; \vm{y}[m], \vm{H}[m] \right) }\IEEEnonumber\\
 &\approx& I^{(\text{a})} \left( \rho, \vm{Q}[m], \vm{R}_{\text{TX}}[m], \vm{Z}[m] \right)\IEEEnonumber\\
 &=& \logdet{\vm{I}_{N_{\text{TX}}} + \rho \vm{R}_{\text{TX}}^{\ast}[m] \vm{Q}[m]} - \frac{\log(e) \rho^2}{2}\IEEEnonumber\\
 &&\times \tr{ \vm{Z}[m] \left( \left( \vm{Q}[m] \left( \vm{I}_{N_{\text{TX}}} + \rho \vm{R}_{\text{TX}}^{\ast}[m] \vm{Q}[m] \right)^{-1} \right)^T \otimes \left( \vm{Q}[m] \left( \vm{I}_{N_{\text{TX}}} + \rho \vm{R}_{\text{TX}}^{\ast}[m] \vm{Q}[m] \right)^{-1} \right) \right) }\IEEEeqnarraynumspace
 \eqnlab{MI_approx}
\end{IEEEeqnarray}
with the $N_{\text{TX}}^2 \times N_{\text{TX}}^2$ fourth-order moment matrix of the channel
\begin{IEEEeqnarray}{rCl}
 \vm{Z}[m] &=& \E{ \vect{\vm{H}^H[m] \vm{H}[m] - \vm{R}_{\text{TX}}^{\ast}[m]} \left( \vect{\vm{H}^H[m] \vm{H}[m] - \vm{R}_{\text{TX}}^{\ast}[m]} \right)^H } .
 \eqnlab{MI_approx_Z}
\end{IEEEeqnarray}

Additionally to $\vm{R}_{\text{TX}}[m]$, \eqnref{MI_approx} requires the evaluation of the fourth-order moment of the channel $\vm{Z}[m]$. In order to gain insight on the influence of typical statistical channel parameters on the MI, we rewrite $\vm{Z}[m]$ as a function of $\bar{\vm{R}}[m]$ and $\tilde{\vm{R}}[m]$ only. Both of these parameters are available with the channel decomposition in \secref{cm_decomposition}. In order to restate \eqnref{MI_approx_Z} for SP as well as for DP channels, we assume that only the co-polarized sub-links can be affected by dominant components. In \appref{MI_approx_cm}, we then obtain the following result:\looseness=-1
\begin{IEEEeqnarray}{rCl}
 \vect{\vm{Z}[m]} &=& \left( \vm{I}_{N_{\text{TX}}} \otimes \vm{Y}[m] \right) \vect{\vm{R}[m]}\IEEEnonumber\\
 &&+ \left( \vm{K}_{N_{\text{TX}},N_{\text{TX}}} \otimes \vm{K}_{N_{\text{TX}},N_{\text{TX}}} \right) \left( \vm{I}_{N_{\text{TX}}} \otimes \vm{Y}^{\ast}[m] \right) \vect{ \bar{\vm{R}}^{\ast}[m] }
 \eqnlab{MI_approx_Z_2}
\end{IEEEeqnarray}
with the $N_{\text{TX}}^3 \times N_{\text{TX}} N_{\text{RX}}^2$ block matrix $\vm{Y}[m]$ containing $\vm{I}_{N_{\text{TX}}} \otimes \vm{X}_{k,l}[m]$ in the $k$th row-partition and the $l$th column-partition for $k = 1, \ldots, N_{\text{TX}}$ and $l = 1, \ldots, N_{\text{RX}}$. The $N_{\text{TX}} \times N_{\text{RX}}$ matrix $\vm{X}_{k,l}[m]$ is defined by $\big[ \vm{X}_{k,l}[m] \big]_{p,q} = \big[ \tilde{\vm{R}}[m] \big]_{(k-1)N_{\text{RX}}+l, (p-1)N_{\text{RX}}+q}$ for $p = 1, \ldots, N_{\text{TX}}$ and $q = 1, \ldots, N_{\text{RX}}$. Note that $\vm{R}[m] = \bar{\vm{R}}[m] + \tilde{\vm{R}}[m]$ holds.

\subsection{SP vs. DP Performance: High-$K$-Factor Case}
\seclab{performance_SP_vs_DP_high_K-factor}

We now compare the performance of SP and DP setups in the high-$K$-factor regime. First, consider the case of an asymptotic $K$-factor setting\ie infinitely large $K$-factors, and that only the co-polarized sub-links have dominant components. Then, the Jensen bound on the MI given by\looseness=-1
\begin{IEEEeqnarray}{rCl}
 I^{(\text{J})} \left( \rho, \vm{Q}[m], \vm{R}_{\text{TX}}[m] \right) &=& \logdet{\vm{I}_{N_{\text{TX}}} + \rho \vm{R}_{\text{TX}}^{\ast}[m] \vm{Q}[m]}
\eqnlab{MI_Jensen}
\end{IEEEeqnarray}
and corresponding to the first term in \eqnref{MI_approx} is equal to the MI \eqnref{MI}; it can thus be used for a simple analytical performance evaluation. Note that the channel influences the Jensen bound on the MI\ie \eqnref{MI_Jensen}, only through $\vm{R}_{\text{TX}}[m]$. In the asymptotic $K$-factor setting, we have $\vm{R}_{\text{TX}}[m] = \bar{\vm{R}}_{\text{TX}}[m]$.

Using Hadamard's inequality \cite[Sec.~7.8.1]{Horn_Matrix_Analysis}, it can be shown that \eqnref{MI_Jensen} is maximized by choosing the eigenvectors of the input covariance matrix $\vm{Q}[m]$ to be given by the eigenvectors of $\vm{R}_{\text{TX}}^{\ast}[m]$. I.e., for the eigendecomposition $\vm{R}_{\text{TX}}^{\ast}[m] = \vm{U}_{\text{TX}}[m] \vm{\Lambda}_{\text{TX}}[m] \vm{U}_{\text{TX}}^H[m]$ with the unitary eigenvector matrix $\vm{U}_{\text{TX}}[m]$ and the diagonal eigenvalue matrix $\vm{\Lambda}_{\text{TX}}[m]$ of $\vm{R}_{\text{TX}}^{\ast}[m]$, we obtain $\vm{Q}[m] = \vm{U}_{\text{TX}}[m] \vm{\Lambda}_{Q}[m] \vm{U}_{\text{TX}}^H[m]$. Here, $\vm{\Lambda}_{Q}[m]$ is the diagonal eigenvalue matrix of $\vm{Q}[m]$ determining the power allocation. Furthermore, we define $\lambda_{\text{TX},k}[m] = [\vm{\Lambda}_{\text{TX}}[m]]_{k,k}$ and $\lambda_{Q,k}[m] = [\vm{\Lambda}_{Q}[m]]_{k,k}$ for $k=1,\ldots,N_{\text{TX}}$, where $\lambda_{\text{TX},k}[m] \geq \lambda_{\text{TX},k+1}[m]$ for $k=1,\ldots,N_{\text{TX}}-1$ holds.

The crossing points between the MI of an SP setup and the MI of a DP setup are then given by\looseness=-1
\begin{IEEEeqnarray}{rCl}
 I^{(\text{J})} \left( \rho, \vm{Q}_{\text{TX,SP}}[m], \vm{R}_{\text{TX,SP}}[m] \right) &=& I^{(\text{J})} \left( \rho, \vm{Q}_{\text{TX,DP}}[m], \vm{R}_{\text{TX,DP}}[m] \right)\\
 \Leftrightarrow \logmod{\prod_{k=1}^{N_{\text{TX}}} \left( 1 + \rho \lambda_{\text{TX,SP},k}[m] \lambda_{Q,\text{SP},k}[m] \right)} &=& \logmod{\prod_{k=1}^{N_{\text{TX}}} \left( 1 + \rho \lambda_{\text{TX,DP},k}[m] \lambda_{Q,\text{DP},k}[m] \right)}
 \eqnlab{MI_Jensen_SP_vs_DP}
\end{IEEEeqnarray}
where $\lambda_{\text{TX,SP},k}[m]$ and $\lambda_{\text{TX,DP},k}[m]$ for $k=1,\ldots,N_{\text{TX}}$ are the eigenvalues of the SP and DP transmit correlation matrices $\vm{R}_{\text{TX,SP}}[m]$ and $\vm{R}_{\text{TX,DP}}[m]$, respectively. Similarly, $\lambda_{Q,\text{SP},k}[m]$ and $\lambda_{Q,\text{DP},k}[m]$ for $k=1,\ldots,N_{\text{TX}}$ are the eigenvalues of the SP and the DP input covariance matrices $\vm{Q}_{\text{TX,SP}}[m]$ and $\vm{Q}_{\text{TX,DP}}[m]$, respectively. As highlighted in \secref{tx_channel_correlation}, we have $\rank{\bar{\vm{R}}_{\text{TX}}[m]} = 2$ if only the co-polarized sub-links have dominant components and we have $\rank{\bar{\vm{R}}_{\text{TX}}[m]} = 1$ in the SP case with a dominant component. In the high-$K$-factor regime with dominant components for co-polarized propagation only, we thus have to decide between an SP setup with one transmitted stream and a DP setup with two transmitted streams.

To obtain the crossing points when $\lambda_{\text{TX,SP},1}[m] > 0$, $\lambda_{\text{TX,DP},k}[m] > 0$, and $\lambda_{Q,\text{DP},k}[m] > 0$ for $k = 1,2$, we simplify \eqnref{MI_Jensen_SP_vs_DP} to
\begin{IEEEeqnarray}{rCl}
 \logmod{1 + \rho \lambda_{\text{TX,SP},1}[m]} &=& \logmod{\prod_{k=1}^2 \left( 1 + \rho \lambda_{\text{TX,DP},k}[m] \lambda_{Q,\text{DP},k}[m]} \right)\\
 \Leftrightarrow 1 + \rho \lambda_{\text{TX,SP},1}[m] &=& \prod_{k=1}^2 \left( 1 + \rho \lambda_{\text{TX,DP},k}[m] \lambda_{Q,\text{DP},k}[m] \right) .
 \eqnlab{MI_Jensen_SP_vs_DP_2}
\end{IEEEeqnarray}
Besides the crossing point at $\rho=0$, there is a crossing point at
\begin{IEEEeqnarray}{rCl}
 \rho_{\text{CP}}^{(\text{J})}[m] = \frac{ \lambda_{\text{TX,SP},1}[m] - \lambda_\text{sum,DP}[m] }{ \lambda_\text{prod,DP}[m] }
 \eqnlab{rho_Jensen_SP_vs_DP}
\end{IEEEeqnarray}
which is positive if $\lambda_{\text{TX,SP},1}[m] > \lambda_\text{sum,DP}[m]$. Here, we defined
\begin{IEEEeqnarray}{rCl}
 \lambda_\text{sum,DP}[m] &=& \lambda_{\text{TX,DP},1}[m] \lambda_{Q,\text{DP},1}[m] + \lambda_{\text{TX,DP},2}[m] \lambda_{Q,\text{DP},2}[m]\\
 \lambda_\text{prod,DP}[m] &=& \lambda_{\text{TX,DP},1}[m] \lambda_{Q,\text{DP},1}[m] \lambda_{\text{TX,DP},2}[m] \lambda_{Q,\text{DP},2}[m] .
 \eqnlab{rho_approx_SP_vs_DP_help}
\end{IEEEeqnarray}
By inspecting \eqnref{MI_Jensen_SP_vs_DP_2}, we observe that the contribution of the MI of the SP setup\ie the left hand side of \eqnref{MI_Jensen_SP_vs_DP_2}, is a linear function of the SNR $\rho$, while the contribution of the MI of the DP setup\ie the right hand side of \eqnref{MI_Jensen_SP_vs_DP_2}, grows quadratically with the SNR $\rho$. We thus conclude that the DP setup outperforms the SP setup only at SNR values above $\rho_{\text{CP}}^{(\text{J})}[m]$ if $\lambda_{\text{TX,SP},1}[m] > \lambda_\text{sum,DP}[m]$ holds. Otherwise, the DP setup always outperforms the SP setup.

\subsection{SP vs. DP Performance: General Case}
\seclab{performance_SP_vs_DP_general}

In this section, we study the performance of the SP and the DP setup in the general case of arbitrary $K$-factors. Now, we need to consider the approximate evaluation of the MI \eqnref{MI_approx} and cannot restrict to the Jensen bound on the MI. Similarly to \secref{performance_SP_vs_DP_high_K-factor}, we consider the case of the SP setup transmitting a single stream and the DP setup transmitting two streams with positive $\lambda_{\text{TX,SP},1}[m]$, $\lambda_{\text{TX,DP},k}[m]$, and $\lambda_{Q,\text{DP},k}[m]$ for $k = 1,2$. Furthermore, we again choose the eigenvectors of $\vm{R}_{\text{TX}}^{\ast}[m]$ as the eigenvectors of the input covariance matrix $\vm{Q}[m]$. In order to get a closed-form expression of the crossing points, we derive a lower bound on the approximate MI \eqnref{MI_approx} in \appref{MI_approx_LB}. It is given by
\begin{IEEEeqnarray}{rCl}
 I^{(\text{LB})} \left( \rho, \vm{Q}[m], \vm{R}_{\text{TX}}[m], \vm{Z}[m] \right) &=& \logdet{\vm{I}_{N_{\text{TX}}} + \rho \vm{R}_{\text{TX}}^{\ast}[m] \vm{Q}[m]} - \log(e) w[m]
 \eqnlab{MI_approx_LB}
\end{IEEEeqnarray}
with
\begin{IEEEeqnarray}{rCl}
 w[m] &=& \sum_{k=1}^{N_{\text{st}}} \sum_{l=1}^{N_{\text{st}}} \frac{ \left[ \big( \vm{U}_{\text{TX}}^T[m] \otimes \vm{U}_{\text{TX}}^H[m] \big) \vm{Z}[m] \big( \vm{U}_{\text{TX}}^{\ast}[m] \otimes \vm{U}_{\text{TX}}[m] \big) \right]_{(k-1)N_{\text{TX}}+l, (k-1)N_{\text{TX}}+l} }{2 \lambda_{\text{TX},k}[m] \lambda_{\text{TX},l}[m]}\IEEEeqnarraynumspace
 \eqnlab{MI_approx_LB_help}
\end{IEEEeqnarray}
and the number of transmitted streams $N_{\text{st}}$. We note that this lower bound is tight in the limit $\rho \rightarrow \infty$. Based on \eqnref{MI_approx_LB}, we calculate the crossing points of the MI of the SP setup and the MI of the DP setup by considering
\begin{IEEEeqnarray}{rCl}
 I^{(\text{LB})} \left( \rho, \vm{Q}_{\text{TX,SP}}[m], \vm{R}_{\text{TX,SP}}[m], \vm{Z}_{\text{SP}}[m] \right) &=& I^{(\text{LB})} \left( \rho, \vm{Q}_{\text{TX,DP}}[m], \vm{R}_{\text{TX,DP}}[m], \vm{Z}_{\text{DP}}[m] \right)
 \eqnlab{MI_approx_SP_vs_DP}
\end{IEEEeqnarray}
where $\vm{Z}_{\text{SP}}[m]$ and $\vm{Z}_{\text{DP}}[m]$ denote the matrix $\vm{Z}[m]$ for the SP and the DP case, respectively. Similar to \secref{performance_SP_vs_DP_high_K-factor}, we note the linear and the quadratic growth with the SNR $\rho$ of the exponentiation (with respect to the base $2$) of the MI \eqnref{MI_approx_LB} for the SP and the DP setup, respectively. We then obtain a crossing point above which the DP setup outperforms the SP setup at
% \begin{IEEEeqnarray}{rCl}
%  \rho_{\text{CP}}^{(\text{LB})}[m] &=& \frac{\lambda_{\text{TX,SP},1}[m] \alpha[m] - \lambda_\text{sum,DP}[m]}{2 \lambda_\text{prod,DP}[m]} \left( 1 + \sqrt{1 - \frac{ 4 (1-\alpha[m]) \lambda_\text{prod,DP}[m] }{ \left( \lambda_{\text{TX,SP},1}[m] \alpha[m] - \lambda_\text{sum,DP}[m] \right)^2 }} \right)\IEEEeqnarraynumspace
%  \eqnlab{rho_approx_SP_vs_DP}
% \end{IEEEeqnarray}
\begin{IEEEeqnarray}{rCl}
 \rho_{\text{CP}}^{(\text{LB})}[m] &=& \frac{\lambda_{\text{TX,SP},1}[m] \alpha[m] - \lambda_\text{sum,DP}[m]}{2 \lambda_\text{prod,DP}[m]} + \sqrt{ \left( \frac{\lambda_{\text{TX,SP},1}[m] \alpha[m] - \lambda_\text{sum,DP}[m]}{2 \lambda_\text{prod,DP}[m]} \right)^2 + \frac{ \alpha[m]-1 }{ \lambda_\text{prod,DP}[m] }}\IEEEnonumber\\
 \eqnlab{rho_approx_SP_vs_DP}
\end{IEEEeqnarray}
if $4 (1-\alpha[m]) \lambda_\text{prod,DP}[m] \leq ( \lambda_{\text{TX,SP},1}[m] \alpha[m] - \lambda_\text{sum,DP}[m] )^2$ is satisfied. Otherwise, the DP setup always outperforms the SP setup. Here, we defined $\alpha[m] = \exp (w_{\text{DP}}[m] - w_{\text{SP}}[m])$, which is a correction factor, and $w_{\text{SP}}[m]$ and $w_{\text{DP}}[m]$ are obtained from \eqnref{MI_approx_LB_help} for the SP and the DP case, respectively. When $\alpha[m] = 1$, we recover the solution \eqnref{rho_Jensen_SP_vs_DP}.

\section{Channel Measurements}
\seclab{channel_measurements}

We evaluate the previously obtained results using urban macrocell channel measurements that were performed at 2.53 GHz in two bands of 45 MHz in Ilmenau, Germany. During the measurement campaign, the DP MIMO channel from three base station (BS) positions with different heights to a multitude of MT tracks was measured sequentially. The MT was moving with a maximal velocity of about $10$~km/h. In this paper, we extract the $20$~MHz band centered at $2.505$~GHz, and we use the three BS positions at a height of $25$~m with the three MT reference tracks. For further details regarding the measurement campaign, see \cite{Schneider_Channel_Reference_Data, Ispas_LQSRegions}.

After denoising the channel measurements in the time-delay domain, we normalize the channel matrices $\vm{H}[m]$. The normalization is performed with a scalar factor such that $\E{||\vm{h}_{\text{co}}[m]||_F^2} = N_{\text{co}}$ is emulated inside each stationarity region containing $N_t=16$ samples in time and $N_f=128$ samples in frequency. Here, $\vm{h}_{\text{co}}[m]$ is a vector containing only the $N_{\text{co}}$ elements of $\vm{H}[m]$ corresponding to co-polarized sub-links. This guarantees a fair comparison between SP and DP setups since we account for the power loss in cross-polarized sub-links. Then, we estimate the statistical quantities by replacing the ensemble averaging with an averaging over $N_t$ time and $N_f$ frequency samples. This yields a total of $2048$ ($\approx 500$ non-coherent) realizations \cite{Ispas_LQSRegions}.

\subsection{Antenna Setups}
\seclab{antenna_setups}

We choose a uniform linear array at the BS and two uniform circular arrays (UCAs), which lie on top of each other, at the MT for the subsequent evaluations. The antenna arrays consist of patch antennas that can be excited vertically and horizontally. Due to the UCAs at the MT, we are able to differentiate between the following four orientations: the front (direction of motion), the back, and the two sides of the MT. For our evaluations, the BS and the MT act as the TX and the RX, respectively. We consider two SP antenna setups, a VP and an HP setup, as well as two DP antenna setups, a co-located (DP-CL) and a spatially separated (DP-SS) setup, for the $4\times4$ MIMO case. For the SP setups, the antennas are separated by $\lambda_c$ at the TX and $0.5 \lambda_c$ (different UCAs) or $0.327 \lambda_c$ (same UCA) at the RX. For the co-located DP-CL setup, the antenna patches at the TX and the RX are separated by $3 \lambda_c$ and $0.5 \lambda_c$ (across the UCAs), respectively. For the spatially separated DP-SS setup, we use the same antenna patches as in the SP case. However, we have a separation of $2 \lambda_c$ between antennas of the same polarization at the TX side. At the RX side, the lower UCA is only used for the VP excitation while the upper UCA is only used for the HP excitation. We note that all setups result in the same array length at the TX.

\subsection{Scenario Classification}
\seclab{classification}

Based on the measurements, for the SP case, we mainly observe links with either low $K$-factors and low correlations between the MIMO sub-links or links with high $K$-factors and high correlations. A similar observation was made in \cite{Jiang_MIMO_DP_Correlation} and \cite{Erceg_MIMO_DP_Capacity}. Thus, similar to \cite{Erceg_MIMO_DP_Capacity}, we classify the measurements into links with low, medium, and high (co-polarized) $K$-factors, see \tabref{reference_links}. The low $K$-factor links are characterized by $K$-factor values in $[0,2]$, while the medium and high $K$-factors links have several peaks with values above $5$ and $10$, respectively. Additionally, we have one link with varying $K$-factors which consists of low and high $K$-factor parts.
\begin{table}[!t]
\centering
\caption{Specification and Properties of the Reference Links}
\vspace{-0.2cm}
\tablab{reference_links}
\begin{tabular}{cccccc} \hline
%  \textbf{Link} & \textbf{BS} & \textbf{Track} & \textbf{MT orient.} & \textbf{MT pos. [m]} & \textbf{$K$-Factors} \\ \hline
\textbf{Link} & \textbf{BS} & \textbf{Track} & \textbf{MT orientation} & \textbf{MT position [m]} & \textbf{$K$-Factors} \\ \hline
 1 & 1 & 41a-42 & back & $0 - 34.9$ & low \\
 2 & 3 & 9a-9b & left & $0 - 38.9$ & medium \\
 3 & 2 & 10b-9a & front & $9.8 - 56.8$ & high \\
 4 & 3 & 10b-9a & left & $0 - 64.9$ & varying \\ \hline
\end{tabular}
\vspace{-0.5cm}
\end{table}
The reason for the low $K$-factors/correlations in link 1 and 2 is that track 41a-42 is partly located in a street canyon; regarding BS 1 and 3 no dominant components are expected. In contrast, tracks 9a-9b and 10b-9a are mostly situated in an open environment where dominant components are more likely to occur.

\section{Results}
\seclab{results}

In order to check the efficiency of the channel decomposition, we compare the $K$-factors from the decomposition to the ones obtained from the measurements with the moment method in \cite{Greenstein_K-Factor_Letter}. The results on the $K$-factors are averaged over the sub-links of each polarization combination for the DP-CL setup. Subsequently, we consider the practically relevant case of extracting $N_{\text{DP}}=2$ eigenvalues, see \secref{cm_decomposition_DP}. In \tabref{Kfactors_4x4}, we show the results for links 1-3 averaged over the driven distance. We see that the cross-polarized sub-links, VP to HP (V-H) and HP to VP (H-V), show significantly smaller $K$-factors than the co-polarized ones, VP to VP (V-V) and HP to HP (H-H). In general, we observe lower $K$-factor values from the channel decomposition; this is due to guaranteeing the positive semidefiniteness of the correlation matrices, which can result in a shift of the power from the dominant components to the remaining components of the channel, see \secref{cm_decomposition}. In \figref{Kfactors_vs_distance_4x4}, we depict the evolution over distance for link 4 since it is characterized by varying $K$-factors, see \tabref{reference_links}. Similar obervations as in \tabref{Kfactors_4x4} can be made. Furthermore, we observe that the channel decomposition is able to reproduce the tendencies in the evolution of the measured $K$-factors.
\begin{table}[!t]
\centering
\caption{Average $K$-Factors From the Measured Channel and the Proposed Channel Decomposition}
\vspace{-0.2cm}
\tablab{Kfactors_4x4}
\begin{tabular}{c|cccc|cccc} \hline
 & \mc{4}{c|}{\textbf{$K$-factors: Measurements}} & \mc{4}{c}{\textbf{$K$-factors: Decomposition}}\\ \hline
 \textbf{Link} & \textbf{V-V} & \textbf{H-H} & \textbf{V-H} & \textbf{H-V} & \textbf{V-V} & \textbf{H-H} & \textbf{V-H} & \textbf{H-V}\\ \hline \hline
 1 & 0.5 & 0.8 & 0.4 & 0.4 & 0.5 & 0.6 & 0.3 & 0.3\\
 2 & 1.6 & 1.4 & 0.6 & 0.7 & 1.2 & 0.9 & 0.2 & 0.2\\
 3 & 4.0 & 5.7 & 1.9 & 1.8 & 4.0 & 5.4 & 1.7 & 1.5\\ \hline
\end{tabular}
% \vspace{-0.5cm}
\end{table}
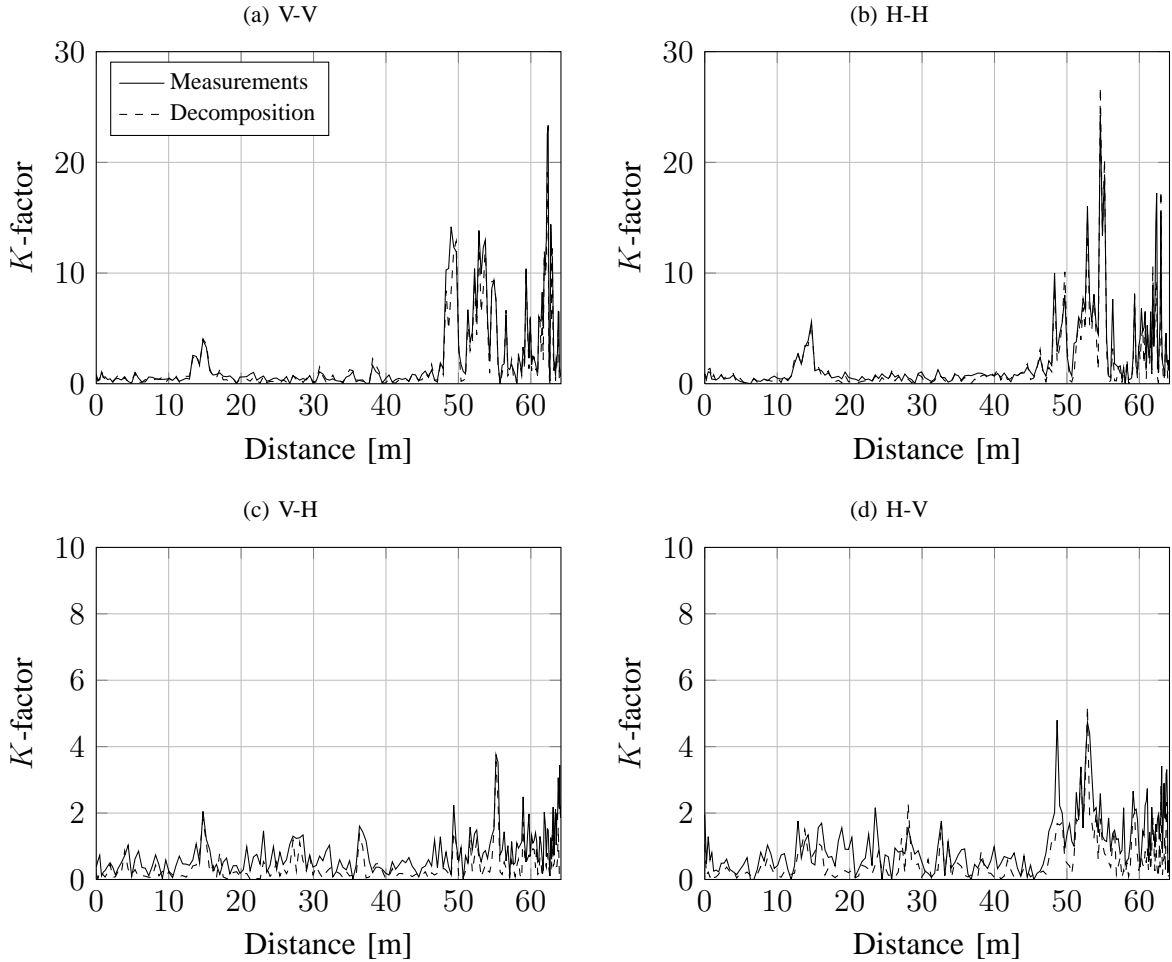
\begin{figure}[!t]
 \centering
 \subfloat[V-V\vspace{-0.2cm}]{
  \begin{tikzpicture}[scale=1]
   \begin{axis}[height=6cm,width=0.47\linewidth,xmin=0,xmax=6.4159333e+01,ymin=0,ymax=30,xlabel={Distance [m]},ylabel={$K$-factor},grid=major,legend pos=north west,legend style={nodes={right},font=\footnotesize},cycle list name=mylist_approx]
    \addplot file {input/K_VV_link6_BS3-MT10b-9a_Tx2V8V2H8H-Rx5Vl5Vu5Hl5Hu_128_161_lower_averaged-16_lseg16_DPmode1.dat};
    \addplot file {input/K_VV_decomp_link6_BS3-MT10b-9a_Tx2V8V2H8H-Rx5Vl5Vu5Hl5Hu_128_161_lower_averaged-16_lseg16_DPmode1.dat};
    \legend{Measurements, Decomposition}
   \end{axis}
  \end{tikzpicture}
 }\hfil
 \subfloat[H-H\vspace{-0.2cm}]{
  \begin{tikzpicture}[scale=1]
   \begin{axis}[height=6cm,width=0.47\linewidth,xmin=0,xmax=6.4159333e+01,ymin=0,ymax=30,xlabel={Distance [m]},ylabel={$K$-factor},grid=major,cycle list name=mylist_approx]
    \addplot file {input/K_HH_link6_BS3-MT10b-9a_Tx2V8V2H8H-Rx5Vl5Vu5Hl5Hu_128_161_lower_averaged-16_lseg16_DPmode1.dat};
    \addplot file {input/K_HH_decomp_link6_BS3-MT10b-9a_Tx2V8V2H8H-Rx5Vl5Vu5Hl5Hu_128_161_lower_averaged-16_lseg16_DPmode1.dat};
   \end{axis}
  \end{tikzpicture}
 }\vspace{-0.2cm}\\
 \subfloat[V-H\vspace{-0.2cm}]{
  \begin{tikzpicture}[scale=1]
   \begin{axis}[height=6cm,width=0.47\linewidth,xmin=0,xmax=6.4159333e+01,ymin=0,ymax=10,xlabel={Distance [m]},ylabel={$K$-factor},grid=major,cycle list name=mylist_approx]
    \addplot file {input/K_VH_link6_BS3-MT10b-9a_Tx2V8V2H8H-Rx5Vl5Vu5Hl5Hu_128_161_lower_averaged-16_lseg16_DPmode1.dat};
    \addplot file {input/K_VH_decomp_link6_BS3-MT10b-9a_Tx2V8V2H8H-Rx5Vl5Vu5Hl5Hu_128_161_lower_averaged-16_lseg16_DPmode1.dat};
   \end{axis}
  \end{tikzpicture}
 }\hfil
 \subfloat[H-V\vspace{-0.2cm}]{
  \begin{tikzpicture}[scale=1]
   \begin{axis}[height=6cm,width=0.47\linewidth,xmin=0,xmax=6.4159333e+01,ymin=0,ymax=10,xlabel={Distance [m]},ylabel={$K$-factor},grid=major,cycle list name=mylist_approx]
    \addplot file {input/K_HV_link6_BS3-MT10b-9a_Tx2V8V2H8H-Rx5Vl5Vu5Hl5Hu_128_161_lower_averaged-16_lseg16_DPmode1.dat};
    \addplot file {input/K_HV_decomp_link6_BS3-MT10b-9a_Tx2V8V2H8H-Rx5Vl5Vu5Hl5Hu_128_161_lower_averaged-16_lseg16_DPmode1.dat};
   \end{axis}
  \end{tikzpicture}
 }
 \vspace{-0.3cm}
 \caption{$K$-factors vs. distance on link 4 (averaged over sub-links of the DP-CL setup with the same polarization combination).}
 \figlab{Kfactors_vs_distance_4x4}
 \vspace{-0.7cm}
\end{figure}

Next, we evaluate the performance of the SP and the DP setups. In order to compare the approximate evaluation of the MI\ie \eqnref{MI_approx} with \eqnref{MI_approx_Z_2}, to the (exact) MI \eqnref{MI}, we use $N_{\text{DP}}=2$. We use the optimal input with respect to the Jensen bound on the MI, where the eigenvectors of $\vm{R}^{\ast}_{\text{TX}}[m]$ form the precoding and the power allocation is obtained by a simple water-filling strategy \cite{Vu_MIMO_Capacity_with_Dynamic_CSIT}, unless otherwise specified. The results of links 1-3 are accumulated over each track and shown as a function of the SNR in \figref{MI_vs_SNR_4x4}. We observe that only at high SNRs there is a noticeable gap between the MI and its approximate evaluation. The DP-CL setup only provides an advantage in terms of the MI compared to the SP setups\ie the VP and the HP setup, if the $K$-factors (of the co-polarized sub-links) and the SNR attain certain values; the higher the $K$-factors, the lower this SNR threshold is. Practically, a switching between SP and DP setups is thus most useful in medium- to high-$K$-factor scenarios; there the crossing points between the MI of the SP setups and the DP-CL setup are accurately reproduced by the approximate evaluation of the MI\ie \eqnref{MI_approx} with \eqnref{MI_approx_Z_2}. Furthermore, in \figref{MI_vs_distance_4x4}, we plot the MI over distance for the VP, the HP, and the DP-CL setup on link 4 at an SNR of $10$~dB. We observe that the positions at which the DP-CL setup outperforms the SP setups coincide with high $K$-factors, see \figref{Kfactors_vs_distance_4x4}.\looseness=-1
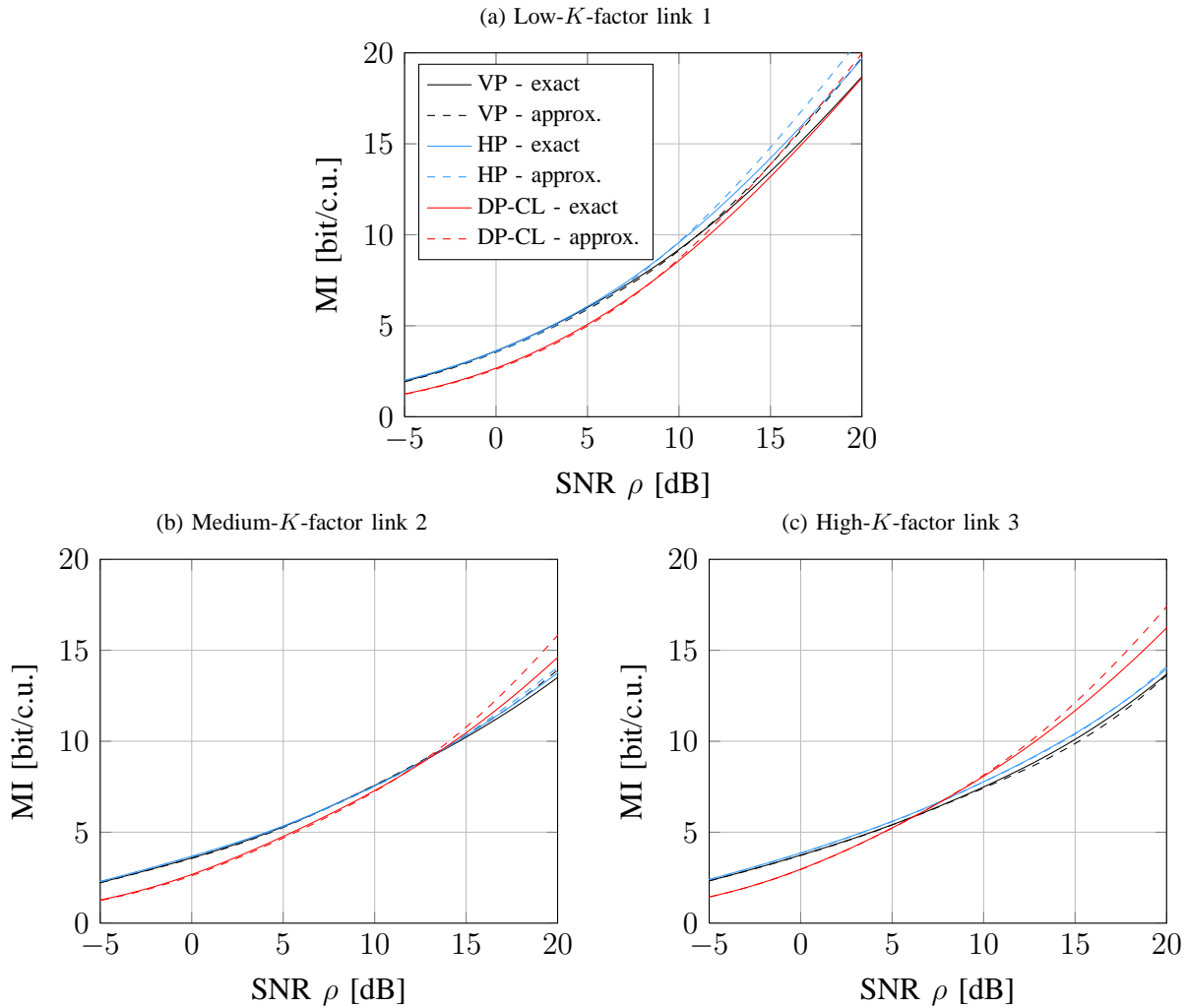
\begin{figure}[!t]
 \centering
 \subfloat[Low-$K$-factor link 1\vspace{-0.2cm}]{
  \begin{tikzpicture}[scale=1]
   \begin{axis}[height=6.5cm,width=0.47\linewidth,xmin=-5,xmax=20,ymin=0,ymax=20,xlabel={SNR $\rho$ [dB]},ylabel={MI [bit/c.u.]},grid=major,legend pos=north west,legend style={nodes={right},font=\footnotesize},cycle list name=mylist_approx]
    \addplot file {input/Cav_Jinput_optimal_link1_BS1-MT41a-42_Tx2V4V6V8V-Rx2Vl2Vu3Vl3Vu_128_161_lower_averaged-16_lseg16_DPmode0.dat};
    \addplot file {input/Cav_Jinput-custom_optimal_link1_BS1-MT41a-42_Tx2V4V6V8V-Rx2Vl2Vu3Vl3Vu_128_161_lower_averaged-16_lseg16_DPmode0.dat};
    \addplot file {input/Cav_Jinput_optimal_link1_BS1-MT41a-42_Tx2H4H6H8H-Rx2Hl2Hu3Hl3Hu_128_161_lower_averaged-16_lseg16_DPmode0.dat};
    \addplot file {input/Cav_Jinput-custom_optimal_link1_BS1-MT41a-42_Tx2H4H6H8H-Rx2Hl2Hu3Hl3Hu_128_161_lower_averaged-16_lseg16_DPmode0.dat};
    \addplot file {input/Cav_Jinput_optimal_link1_BS1-MT41a-42_Tx2V8V2H8H-Rx2Vl2Vu2Hl2Hu_128_161_lower_averaged-16_lseg16_DPmode1.dat};
    \addplot file {input/Cav_Jinput-custom_optimal_link1_BS1-MT41a-42_Tx2V8V2H8H-Rx2Vl2Vu2Hl2Hu_128_161_lower_averaged-16_lseg16_DPmode1.dat};
    \legend{VP - exact, VP - approx., HP - exact, HP - approx., DP-CL - exact, DP-CL - approx.}
   \end{axis}
  \end{tikzpicture}
 }\vspace{-0.5cm}\\
 \subfloat[Medium-$K$-factor link 2\vspace{-0.2cm}]{
  \begin{tikzpicture}[scale=1]
   \begin{axis}[height=6.5cm,width=0.47\linewidth,xmin=-5,xmax=20,ymin=0,ymax=20,xlabel={SNR $\rho$ [dB]},ylabel={MI [bit/c.u.]},grid=major,cycle list name=mylist_approx]
    \addplot file {input/Cav_Jinput_optimal_link3_BS3-MT9a-9b_Tx2V4V6V8V-Rx5Vl5Vu6Vl6Vu_128_161_lower_averaged-16_lseg16_DPmode0.dat};
    \addplot file {input/Cav_Jinput-custom_optimal_link3_BS3-MT9a-9b_Tx2V4V6V8V-Rx5Vl5Vu6Vl6Vu_128_161_lower_averaged-16_lseg16_DPmode0.dat};
    \addplot file {input/Cav_Jinput_optimal_link3_BS3-MT9a-9b_Tx2H4H6H8H-Rx5Hl5Hu6Hl6Hu_128_161_lower_averaged-16_lseg16_DPmode0.dat};
    \addplot file {input/Cav_Jinput-custom_optimal_link3_BS3-MT9a-9b_Tx2H4H6H8H-Rx5Hl5Hu6Hl6Hu_128_161_lower_averaged-16_lseg16_DPmode0.dat};
    \addplot file {input/Cav_Jinput_optimal_link3_BS3-MT9a-9b_Tx2V8V2H8H-Rx5Vl5Vu5Hl5Hu_128_161_lower_averaged-16_lseg16_DPmode1.dat};
    \addplot file {input/Cav_Jinput-custom_optimal_link3_BS3-MT9a-9b_Tx2V8V2H8H-Rx5Vl5Vu5Hl5Hu_128_161_lower_averaged-16_lseg16_DPmode1.dat};
   \end{axis}
  \end{tikzpicture}
 }\hfil
 \subfloat[High-$K$-factor link 3\vspace{-0.2cm}]{
  \begin{tikzpicture}[scale=1]
   \begin{axis}[height=6.5cm,width=0.47\linewidth,xmin=-5,xmax=20,ymin=0,ymax=20,xlabel={SNR $\rho$ [dB]},ylabel={MI [bit/c.u.]},grid=major,cycle list name=mylist_approx]
    \addplot file {input/Cav_Jinput_optimal_link5_BS2-MT10b-9a_Tx2V4V6V8V-Rx8Vl8Vu9Vl9Vu_128_161_lower_averaged-16_lseg16_DPmode0.dat};
    \addplot file {input/Cav_Jinput-custom_optimal_link5_BS2-MT10b-9a_Tx2V4V6V8V-Rx8Vl8Vu9Vl9Vu_128_161_lower_averaged-16_lseg16_DPmode0.dat};
    \addplot file {input/Cav_Jinput_optimal_link5_BS2-MT10b-9a_Tx2H4H6H8H-Rx8Hl8Hu9Hl9Hu_128_161_lower_averaged-16_lseg16_DPmode0.dat};
    \addplot file {input/Cav_Jinput-custom_optimal_link5_BS2-MT10b-9a_Tx2H4H6H8H-Rx8Hl8Hu9Hl9Hu_128_161_lower_averaged-16_lseg16_DPmode0.dat};
    \addplot file {input/Cav_Jinput_optimal_link5_BS2-MT10b-9a_Tx2V8V2H8H-Rx8Vl8Vu8Hl8Hu_128_161_lower_averaged-16_lseg16_DPmode1.dat};
    \addplot file {input/Cav_Jinput-custom_optimal_link5_BS2-MT10b-9a_Tx2V8V2H8H-Rx8Vl8Vu8Hl8Hu_128_161_lower_averaged-16_lseg16_DPmode1.dat};
   \end{axis}
  \end{tikzpicture}
 }
 \caption{MI vs. SNR of the exact and the approximate evaluation for the VP, the HP, and the DP-CL setup.}
 \figlab{MI_vs_SNR_4x4}
 \vspace{-0.5cm}
\end{figure}
\begin{figure}[!t]
 \centering
 \begin{tikzpicture}[scale=1]
  \begin{axis}[height=5.5cm,width=\columnwidth,xmin=0,xmax=6.4159333e+01,ymin=0,ymax=10,xlabel={Distance [m]},ylabel={MI [bit/c.u.]},grid=major,legend pos=south west,legend style={nodes={right},font=\footnotesize},cycle list name=mylist_approx]
   \addplot[opacity=0.4,forget plot,draw=none,fill=blue!30!white] coordinates{(12,10) (17,10)}\closedcycle;
   \addplot[opacity=0.4,forget plot,draw=none,fill=blue!30!white] coordinates{(47,10) (70,10)}\closedcycle;
   \addplot file {input/C_Jinput_optimal_SNR10_link6_BS3-MT10b-9a_Tx2V4V6V8V-Rx5Vl5Vu6Vl6Vu_128_161_lower_averaged-16_lseg16_DPmode0.dat};
   \addplot file {input/C_Jinput-custom_optimal_SNR10_link6_BS3-MT10b-9a_Tx2V4V6V8V-Rx5Vl5Vu6Vl6Vu_128_161_lower_averaged-16_lseg16_DPmode0.dat};
   \addplot file {input/C_Jinput_optimal_SNR10_link6_BS3-MT10b-9a_Tx2H4H6H8H-Rx5Hl5Hu6Hl6Hu_128_161_lower_averaged-16_lseg16_DPmode0.dat};
   \addplot file {input/C_Jinput-custom_optimal_SNR10_link6_BS3-MT10b-9a_Tx2H4H6H8H-Rx5Hl5Hu6Hl6Hu_128_161_lower_averaged-16_lseg16_DPmode0.dat};
   \addplot file {input/C_Jinput_optimal_SNR10_link6_BS3-MT10b-9a_Tx2V8V2H8H-Rx5Vl5Vu5Hl5Hu_128_161_lower_averaged-16_lseg16_DPmode1.dat};
   \addplot file {input/C_Jinput-custom_optimal_SNR10_link6_BS3-MT10b-9a_Tx2V8V2H8H-Rx5Vl5Vu5Hl5Hu_128_161_lower_averaged-16_lseg16_DPmode1.dat};
   \legend{VP - exact, VP - approx., HP - exact, HP - approx., DP-CL - exact, DP-CL - approx.}
  \end{axis}
 \end{tikzpicture}
 \vspace{-0.3cm}
 \caption{MI vs. distance for the exact and the approximate evaluation for the VP, the HP, and the DP-CL setup on link 4 with an SNR $\rho=10$~dB (the blue-shaded regions denote positions where the (co-polarized) $K$-factors are high, cf. \figref{Kfactors_vs_distance_4x4}).}
 \figlab{MI_vs_distance_4x4}
%  \vspace{-0.5cm}
\end{figure}
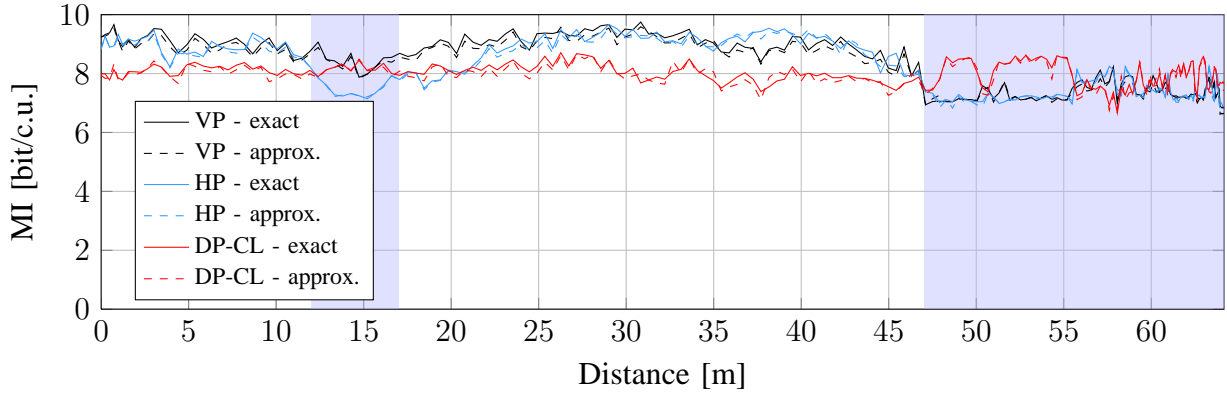

We now compare the performance using two different DP setups, the DP-CL setup with co-located antennas and the DP-SS setup with spatially separated antennas. In \figref{MI_vs_SNR_DP_link_3_4x4}, we show the MI of the DP-CL and the DP-SS setup, exemplarily, on link 2. We observe that the DP-SS setup is able to reach even higher MI values at high SNR. We expect that this is due to the increased viewing angle into the propagation channel for each polarization at the RX side, which results in an increase in the degrees of freedom. The DP-CL setup, however, offers a more compact antenna array at the cost of a reduced viewing angle at the RX. Furthermore, we observe here that the approximate evaluation of the MI is more accurate for the DP-SS setup than it is for the DP-CL setup.
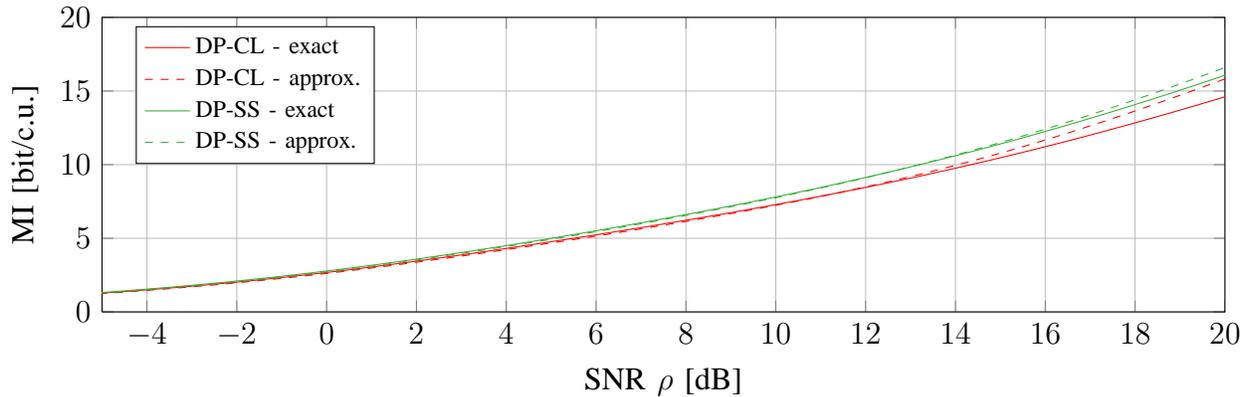
\begin{figure}[!t]
 \centering
 \begin{tikzpicture}[scale=1]
  \begin{axis}[height=5.5cm,width=\columnwidth,xmin=-5,xmax=20,ymin=0,ymax=20,xlabel={SNR $\rho$ [dB]},ylabel={MI [bit/c.u.]},grid=major,legend pos=north west,legend style={nodes={right},font=\footnotesize},cycle list name=mylist_DP_approx]
   \addplot file {input/Cav_Jinput_optimal_link3_BS3-MT9a-9b_Tx2V8V2H8H-Rx5Vl5Vu5Hl5Hu_128_161_lower_averaged-16_lseg16_DPmode1.dat};
   \addplot file {input/Cav_Jinput-custom_optimal_link3_BS3-MT9a-9b_Tx2V8V2H8H-Rx5Vl5Vu5Hl5Hu_128_161_lower_averaged-16_lseg16_DPmode1.dat};
   \addplot file {input/Cav_Jinput_optimal_link3_BS3-MT9a-9b_Tx2V6V4H8H-Rx5Vl6Vl5Hu6Hu_128_161_lower_averaged-16_lseg16_DPmode1.dat};
   \addplot file {input/Cav_Jinput-custom_optimal_link3_BS3-MT9a-9b_Tx2V6V4H8H-Rx5Vl6Vl5Hu6Hu_128_161_lower_averaged-16_lseg16_DPmode1.dat};
   \legend{DP-CL - exact, DP-CL - approx., DP-SS - exact, DP-SS - approx.}
  \end{axis}
 \end{tikzpicture}
 \vspace{-1.3cm}
 \caption{MI vs. SNR for the DP-CL and the DP-SS setup on link 2.}
 \figlab{MI_vs_SNR_DP_link_3_4x4}
%  \vspace{-0.3cm}
\end{figure}

The average SNR values above which the MI of the DP-CL setup with two streams and equal power allocation is higher than the MI of the VP or the HP setup with a single stream are given \tabref{CP_4x4}. Note that the precoding is again given by the eigenvectors of $\vm{R}^{\ast}_{\text{TX}}[m]$. The resulting crossing points are calculated using the various methods introduced before\ie using the MI and the approximations given in \eqnref{MI}, \eqnref{MI_approx}, \eqnref{MI_Jensen}, and \eqnref{MI_approx_LB} together with \eqnref{MI_approx_Z_2}. We observe that the approximate evaluation of the MI \eqnref{MI_approx} is able to accurately reproduce the average SNR values. When using the Jensen bound on the MI, we obtain lower average SNR values. Note that the Jensen bound on the MI is only useful for high-$K$-factor links; thus, we only give the results for link 3. The SNR values obtained from the lower bound on the approximate MI\ie \eqnref{MI_approx_LB}, yield a slight overestimation of the average SNR values for all links. We observe that all the (exact) crossing points are roughly between $5$ and $7$~dB. A clear dependence on the link is not present; this is due to the restriction to two and one transmitted stream for DP and SP MIMO systems, respectively.
\begin{table}[!t]
\centering
\caption{Average SNR Values Above Which the MI of the DP-CL Setup with Two Streams and Equal Power Allocation is Higher than the MI of an SP Setup with a Single Stream}
\tablab{CP_4x4}
\begin{tabular}{c|ccc|ccc} \hline
& \mc{6}{c}{\textbf{SNR Values $\rho_{\text{CP}}$ [dB] (averaged)}}\\ \hline
& \mc{3}{c|}{\textbf{VP vs. DP-CL}} & \mc{3}{c}{\textbf{HP vs. DP-CL}}\\ \hline
 \textbf{Method} & \textbf{Link 1} & \textbf{Link 2} & \textbf{Link 3} & \textbf{Link 1} & \textbf{Link 2} & \textbf{Link 3}\\ \hline \hline
\hspace{-0.2cm}Exact: \eqnref{MI}\hspace{-0.1cm} & 4.998 & 6.759 & 5.242   & 5.559 & 7.197 & 5.744\\
\hspace{-0.2cm}Appr.: \eqnref{MI_approx}\hspace{-0.1cm} & 5.073 & 7.027 & 5.240   & 5.567 & 7.324 & 5.680\\
\hspace{-0.2cm}$\rho_{\text{CP}}^{(\text{J})}$: \eqnref{MI_Jensen}\hspace{-0.1cm} & --- & --- & 4.623   & --- & --- & 4.813\\
\hspace{-0.2cm}$\rho_{\text{CP}}^{(\text{LB})}$: \eqnref{MI_approx_LB}\hspace{-0.1cm} & 6.154 & 7.722 & 5.747   & 6.564 & 7.976 & 6.130\\
\end{tabular}
% \vspace{-0.5cm}
\end{table}

\section{Conclusion}
\seclab{conclusion}

In this paper, we have studied the modeling of DP MIMO channels as well as the performance over such channels. We proposed a general model for DP mobile Ricean channels with a channel decomposition technique yielding necessary statistical channel parameters. Furthermore, we derived an approximation of the MI, which is a function of those parameters, in order to gain some understanding on the statistical channel parameters influencing the MI. Based on the approximate evaluation of the MI, we were able to analytically characterize the required SNR for a dual-stream DP MIMO system to outperform a single-stream SP MIMO system. Finally, we applied the obtained results to channel measurements performed in an urban macrocell environment at $2.53$~GHz. We find that for sufficiently high $K$-factors DP MIMO systems are able to outperform SP MIMO systems if a certain, practically relevant, SNR is attained.

% \section*{Acknowledgment}

% \appendix
\appendices

\section{Rank of $\bar{\vm{R}}[m]$}
\applab{R_bar_rank}

We are interested in a condition on the rank of $\bar{\vm{R}}[m]$ for the DP case. We first drop the time argument for notational simplicity. Then, we rearrange $\bar{\vm{R}}$ through column and row permutations with the permutation matrix $\vm{P}$ into $\bar{\vm{R}}^{(\text{p})} = \vm{P} \bar{\vm{R}} \vm{P}^T$ such that
\begin{IEEEeqnarray}{rCl}
 \bar{\vm{R}}^{(\text{p})} &=& \begin{bmatrix}
 \bar{\vm{R}}_{\text{VVVV}} & \bar{\vm{R}}_{\text{VVVH}} & \bar{\vm{R}}_{\text{VVHV}} & \bar{\vm{R}}_{\text{VVHH}}\\
 \bar{\vm{R}}_{\text{VHVV}} & \bar{\vm{R}}_{\text{VHVH}} & \bar{\vm{R}}_{\text{VHHV}} & \bar{\vm{R}}_{\text{VHHH}}\\
 \bar{\vm{R}}_{\text{HVVV}} & \bar{\vm{R}}_{\text{HVVH}} & \bar{\vm{R}}_{\text{HVHV}} & \bar{\vm{R}}_{\text{HVHH}}\\
 \bar{\vm{R}}_{\text{HHVV}} & \bar{\vm{R}}_{\text{HHVH}} & \bar{\vm{R}}_{\text{HHHV}} & \bar{\vm{R}}_{\text{HHHH}}
 \end{bmatrix}
 \eqnlab{corr_matrix_perm_definition}
\end{IEEEeqnarray}
with $\bar{\vm{R}}_{abcd} = \E{ \vect{\bar{\vm{H}}_{ab}} \left( \vect{\bar{\vm{H}}_{cd}} \right)^H }$ for $a,b,c,d \in \{ \text{V},\text{H} \}$ holds. We now have
% \begin{IEEEeqnarray}{rCl}
%  \bar{\vm{R}}^{(\text{p})} &\stackrel{(\text{a})}{=}& \left( \vect{\vm{V}} \left( \vect{\vm{V}} \right)^H \right) \odot \E{\vect{\vm{\Phi}} \left( \vect{\vm{\Phi}} \right)^H}\IEEEnonumber\\
%  &\stackrel{(b)}{=}& \left( \vect{\vm{V}} \left( \vect{\vm{V}} \right)^H \right)\IEEEnonumber\\
%  && \odot \left( \vect{\vm{\Delta}_{\phi}} \left( \vect{\vm{\Delta}_{\phi} \right)^H} \right)\IEEEnonumber\\
%  && \odot \left( \vm{G} \otimes \vm{1}_{\frac{N_{\text{TX}} N_{\text{RX}}}{4}} \right)
%  \eqnlab{corr_matrix_perm_relation}
% \end{IEEEeqnarray}
\begin{IEEEeqnarray}{rCl}
 \bar{\vm{R}}^{(\text{p})} &\stackrel{(\text{a})}{=}& \left( \vect{\vm{V}} \left( \vect{\vm{V}} \right)^H \right) \odot \E{\vect{\vm{\Phi}} \left( \vect{\vm{\Phi}} \right)^H}\IEEEnonumber\\
 &\stackrel{(b)}{=}& \left( \vect{\vm{V}} \left( \vect{\vm{V}} \right)^H \right) \odot \left( \vect{\vm{\Delta}_{\phi}} \left( \vect{\vm{\Delta}_{\phi}} \right)^H \right) \odot \left( \vm{G} \otimes \vm{1}_{\frac{N_{\text{TX}} N_{\text{RX}}}{4}} \right)
 \eqnlab{corr_matrix_perm_relation}
\end{IEEEeqnarray}
where, in (a), we used \eqnref{channel_model_dominant} and defined
% \begin{IEEEeqnarray}{rCl}
%  \vect{\vm{V}} &=& \left[ \left( \vect{\vm{V}_{\text{VV}}} \right)^T \left( \vect{\vm{V}_{\text{VH}}} \right)^T \right.\IEEEnonumber\\
%  && \left. \left( \vect{\vm{V}_{\text{HV}}} \right)^T \left( \vect{\vm{V}_{\text{HH}}} \right)^T \right]^T\\
%  \vect{\vm{\Phi}} &=& \left[ \left( \vect{\vm{\Phi}_{\text{VV}}} \right)^T \left( \vect{\vm{\Phi}_{\text{VH}}} \right)^T \right.\IEEEnonumber\\
%  && \left. \left( \vect{\vm{\Phi}_{\text{HV}}} \right)^T \left( \vect{\vm{\Phi}_{\text{HH}}} \right)^T \right]^T
% \end{IEEEeqnarray}
\begin{IEEEeqnarray}{rCl}
 \vect{\vm{V}} &=& \left[ \left( \vect{\vm{V}_{\text{VV}}} \right)^T \left( \vect{\vm{V}_{\text{VH}}} \right)^T \left( \vect{\vm{V}_{\text{HV}}} \right)^T \left( \vect{\vm{V}_{\text{HH}}} \right)^T \right]^T\\
 \vect{\vm{\Phi}} &=& \left[ \left( \vect{\vm{\Phi}_{\text{VV}}} \right)^T \left( \vect{\vm{\Phi}_{\text{VH}}} \right)^T \left( \vect{\vm{\Phi}_{\text{HV}}} \right)^T \left( \vect{\vm{\Phi}_{\text{HH}}} \right)^T \right]^T
\end{IEEEeqnarray}
and, in (b), we used \eqnref{channel_model_dominant_rewrite} and defined
% \begin{IEEEeqnarray}{rCl}
%  \vect{\vm{\Delta}_{\phi}} &=& \left[ \left( \vect{\vm{\Delta}_{\phi,\text{VV}}} \right)^T \left( \vect{\vm{\Delta}_{\phi,\text{VH}}} \right)^T \right.\IEEEnonumber\\
%  && \left. \left( \vect{\vm{\Delta}_{\phi,\text{HV}}} \right)^T \left( \vect{\vm{\Delta}_{\phi,\text{HH}}} \right)^T \right]^T\\
%  \vm{G} &=& \begin{bmatrix}
%  1 & g_{\text{VVVH}} & g_{\text{VVHV}} & g_{\text{VVHH}}\\
%  g_{\text{VHVV}} & 1 & g_{\text{VHHV}} & g_{\text{VHHH}}\\
%  g_{\text{HVVV}} & g_{\text{HVVH}} & 1 & g_{\text{HVHH}}\\
%  g_{\text{HHVV}} & g_{\text{HHVH}} & g_{\text{HHHV}} & 1
%  \end{bmatrix}
% \end{IEEEeqnarray}
% \begin{IEEEeqnarray}{rCl}
%  \vect{\vm{\Delta}_{\phi}} &=& \left[ \left( \vect{\vm{\Delta}_{\phi,\text{VV}}} \right)^T \left( \vect{\vm{\Delta}_{\phi,\text{VH}}} \right)^T \left( \vect{\vm{\Delta}_{\phi,\text{HV}}} \right)^T \left( \vect{\vm{\Delta}_{\phi,\text{HH}}} \right)^T \right]^T\\
%  \vm{G} &=& \begin{bmatrix}
%  1 & g_{\text{VVVH}} & g_{\text{VVHV}} & g_{\text{VVHH}}\\
%  g_{\text{VHVV}} & 1 & g_{\text{VHHV}} & g_{\text{VHHH}}\\
%  g_{\text{HVVV}} & g_{\text{HVVH}} & 1 & g_{\text{HVHH}}\\
%  g_{\text{HHVV}} & g_{\text{HHVH}} & g_{\text{HHHV}} & 1
%  \end{bmatrix}
% \end{IEEEeqnarray}
\begin{IEEEeqnarray}{rCl}
 \vect{\vm{\Delta}_{\phi}} &=& \left[ \left( \vect{\vm{\Delta}_{\phi,\text{VV}}} \right)^T \left( \vect{\vm{\Delta}_{\phi,\text{VH}}} \right)^T \left( \vect{\vm{\Delta}_{\phi,\text{HV}}} \right)^T \left( \vect{\vm{\Delta}_{\phi,\text{HH}}} \right)^T \right]^T\\
 \vm{G} &=& \left[ \begin{IEEEeqnarraybox*}[][c]{,c/c/c/c,}
 1 & g_{\text{VVVH}} & g_{\text{VVHV}} & g_{\text{VVHH}}\\
 g_{\text{VHVV}} & 1 & g_{\text{VHHV}} & g_{\text{VHHH}}\\
 g_{\text{HVVV}} & g_{\text{HVVH}} & 1 & g_{\text{HVHH}}\\
 g_{\text{HHVV}} & g_{\text{HHVH}} & g_{\text{HHHV}} & 1
 \end{IEEEeqnarraybox*} \right]
\end{IEEEeqnarray}
with $g_{abcd} = \E{e^{j(\phi_{ab}-\phi_{cd})}}$ for $a,b,c,d \in \{ \text{V},\text{H} \}$ and the all-one matrix $\vm{1}_{N}$ of size $N \times N$. As the rank of a matrix is unchanged by left or right multiplication with a non-singular matrix \cite[Sec.~0.4.6~(b)]{Horn_Matrix_Analysis}, it is obvious that
%  $\rank{\bar{\vm{R}}} = \rank{\bar{\vm{R}}^{(\text{p})}}$ and $\rank{\vm{G}} \leq 4$ 
\begin{IEEEeqnarray}{rCl}
 \rank{\bar{\vm{R}}} &=& \rank{\bar{\vm{R}}^{(\text{p})}}
 \eqnlab{app_corr_matrix_rank_relation}\\
 \rank{\vm{G}} &\leq& 4
 \eqnlab{app_phase_rank_relation}
\end{IEEEeqnarray}
hold. Moreover, we have $\rank{\vm{A} \odot \vm{B}} \leq \rank{\vm{A}} \rank{\vm{B}}$ \cite[Th.~5.1.7]{Horn_Matrix_Analysis_2} as well as $\rank{\vm{A} \otimes \vm{B}} = \rank{\vm{A}} \rank{\vm{B}}$ \cite[Th.~4.2.15]{Horn_Matrix_Analysis_2} for matrices $\vm{A}$ and $\vm{B}$ of appropriate sizes. 
% With \eqnref{corr_matrix_perm_relation}, we then immediately obtain the inequality
With \eqnref{corr_matrix_perm_relation}, \eqnref{app_corr_matrix_rank_relation}, and \eqnref{app_phase_rank_relation}, we then immediately obtain the inequality
\begin{IEEEeqnarray}{rCl}
 \rank{\bar{\vm{R}}} &\leq& 4 .
 \eqnlab{app_corr_matrix_rank_inequality}
\end{IEEEeqnarray}

\section{Rank of $\bar{\vm{R}}_{\text{TX}}[m]$}
\applab{R_tx_bar_rank}

We first drop the time argument to simplify notation. In order to evaluate the rank of $\bar{\vm{R}}_{\text{TX}}$, we use $\vm{V}_{ab} = \vm{v}_{\text{RX},ab} \vm{v}_{\text{TX},ab}^T$ and $\vm{\Delta}_{\vm{\phi},ab} = \vm{d}_{\text{RX},ab} \vm{d}_{\text{TX},ab}^T$. Based on \eqnref{channel_model_dominant_rewrite}, we decompose the dominant channel component for each polarization combination $a,b \in \{ \text{V},\text{H} \}$ as
\begin{IEEEeqnarray}{rCl}
 \bar{\vm{H}}_{ab} &=& \left( \vm{v}_{\text{RX},ab} \odot \vm{d}_{\text{RX},ab} \right) \left( \vm{v}_{\text{TX},ab} \odot \vm{d}_{\text{TX},ab} \right)^T ~e^{j \phi_{ab}} .
 \eqnlab{app_channel_model_dominant_rewrite_2}
\end{IEEEeqnarray}
We then obtain for $a,b,c,d \in \{ \text{V},\text{H} \}$
\begin{IEEEeqnarray}{rCl}
 \E{ \bar{\vm{H}}_{ab}^T \bar{\vm{H}}_{cd}^{\ast} } &=& \left( \vm{v}_{\text{TX},ab} \odot \vm{d}_{\text{TX},ab} \right) \left( \vm{v}_{\text{TX},cd} \odot \vm{d}_{\text{TX},cd} \right)^H f_{abcd}
 \eqnlab{app_TX_corr_matrix_dom_part}
\end{IEEEeqnarray}
with $f_{abcd} = \left( \vm{v}_{\text{RX},ab} \odot \vm{d}_{\text{RX},ab} \right)^T \left( \vm{v}_{\text{RX},cd} \odot \vm{d}_{\text{RX},cd} \right)^{\ast} \E{e^{j (\phi_{ab}-\phi_{cd}) }}$. With \eqnref{channel_model}, we can write
\begin{IEEEeqnarray}{rCl}
 \bar{\vm{R}}_{\text{TX}} &=& \E{ \begin{bmatrix} \bar{\vm{H}}_{\text{VV}}^T \bar{\vm{H}}_{\text{VV}}^{\ast} + \bar{\vm{H}}_{\text{VH}}^T \bar{\vm{H}}_{\text{VH}}^{\ast}, & \bar{\vm{H}}_{\text{VV}}^T \bar{\vm{H}}_{\text{HV}}^{\ast} + \bar{\vm{H}}_{\text{VH}}^T \bar{\vm{H}}_{\text{HH}}^{\ast}\\
 \bar{\vm{H}}_{\text{HV}}^T \bar{\vm{H}}_{\text{VV}}^{\ast} + \bar{\vm{H}}_{\text{HH}}^T \bar{\vm{H}}_{\text{VH}}^{\ast}, & \bar{\vm{H}}_{\text{HV}}^T \bar{\vm{H}}_{\text{HV}}^{\ast} + \bar{\vm{H}}_{\text{HH}}^T \bar{\vm{H}}_{\text{HH}}^{\ast} \end{bmatrix} }
 \eqnlab{app_TX_corr_matrix_dom}
\end{IEEEeqnarray}
Using \eqnref{app_TX_corr_matrix_dom} with \eqnref{app_TX_corr_matrix_dom_part}, we obtain
% \begin{IEEEeqnarray}{rCl}
%  \bar{\vm{R}}_{\text{TX}} &=& \left[ \begin{bmatrix} \left( \vm{v}_{\text{TX},\text{VV}} \odot \vm{d}_{\text{TX},\text{VV}} \right) f_{\text{VVVV}}\\ \left( \vm{v}_{\text{TX},\text{HV}} \odot \vm{d}_{\text{TX},\text{HV}} \right) f_{\text{HVVV}} \end{bmatrix} \left( \vm{v}_{\text{TX},\text{VV}} \odot \vm{d}_{\text{TX},\text{VV}} \right)^H\right.\IEEEnonumber\\
%  &&\left.~\begin{bmatrix} \left( \vm{v}_{\text{TX},\text{VV}} \odot \vm{d}_{\text{TX},\text{VV}} \right) f_{\text{VVHV}}\\ \left( \vm{v}_{\text{TX},\text{HV}} \odot \vm{d}_{\text{TX},\text{HV}} \right) f_{\text{HVHV}} \end{bmatrix} \left( \vm{v}_{\text{TX},\text{HV}} \odot \vm{d}_{\text{TX},\text{HV}} \right)^H \right]\IEEEnonumber\\
%  &&+\left[ \begin{bmatrix} \left( \vm{v}_{\text{TX},\text{VH}} \odot \vm{d}_{\text{TX},\text{VH}} \right) f_{\text{VHVH}}\\ \left( \vm{v}_{\text{TX},\text{HH}} \odot \vm{d}_{\text{TX},\text{HH}} \right) f_{\text{HHVH}} \end{bmatrix} \left( \vm{v}_{\text{TX},\text{VH}} \odot \vm{d}_{\text{TX},\text{VH}} \right)^H\right.\IEEEnonumber\\
%  &&\left.~~~~\begin{bmatrix} \left( \vm{v}_{\text{TX},\text{VH}} \odot \vm{d}_{\text{TX},\text{VH}} \right) f_{\text{VHHH}}\\ \left( \vm{v}_{\text{TX},\text{HH}} \odot \vm{d}_{\text{TX},\text{HH}} \right) f_{\text{HHHH}} \end{bmatrix} \left( \vm{v}_{\text{TX},\text{HH}} \odot \vm{d}_{\text{TX},\text{HH}} \right)^H \right] .
%  \eqnlab{app_TX_corr_matrix_dom_2}
% \end{IEEEeqnarray}
\begin{IEEEeqnarray}{rCl}
 \bar{\vm{R}}_{\text{TX}} &=& \left[ \begin{bmatrix} \vm{t}_{\text{VV}} f_{\text{VVVV}}\\ \vm{t}_{\text{HV}} f_{\text{HVVV}} \end{bmatrix} \vm{t}_{\text{VV}}^H,~ \begin{bmatrix} \vm{t}_{\text{VV}} f_{\text{VVHV}}\\ \vm{t}_{\text{HV}} f_{\text{HVHV}} \end{bmatrix} \vm{t}_{\text{HV}}^H \right] + \left[ \begin{bmatrix} \vm{t}_{\text{VH}} f_{\text{VHVH}}\\ \vm{t}_{\text{HH}} f_{\text{HHVH}} \end{bmatrix} \vm{t}_{\text{VH}}^H,~ \begin{bmatrix} \vm{t}_{\text{VH}} f_{\text{VHHH}}\\ \vm{t}_{\text{HH}} f_{\text{HHHH}} \end{bmatrix} \vm{t}_{\text{HH}}^H \right]\IEEEeqnarraynumspace
 \eqnlab{app_TX_corr_matrix_dom_2}
\end{IEEEeqnarray}
with $\vm{t}_{ab} = \vm{v}_{\text{TX},ab} \odot \vm{d}_{\text{TX},ab}$ for $a,b \in \{ \text{V},\text{H} \}$. For matrices $\vm{A}$ and $\vm{B}$ of appropriate sizes, we have $\rank{\vm{A} + \vm{B}} \leq \rank{\vm{A}} + \rank{\vm{B}}$ \cite[Sec.~0.4.5~(d)]{Horn_Matrix_Analysis_2}. Thus, we conclude that
\begin{IEEEeqnarray}{rCl}
 \rank{\bar{\vm{R}}_{\text{TX}}} &\leq& 4
 \eqnlab{app_tx_corr_matrix_rank_inequality}
\end{IEEEeqnarray}
must hold. If only the co-polarized sub-links have dominant components, $\rank{\bar{\vm{R}}_{\text{TX}}} = 2$ is obtained using \eqnref{app_TX_corr_matrix_dom_2}. For an SP setup with a dominant component, we have $\rank{\bar{\vm{R}}_{\text{TX}}} = 1$.

\section{Evaluation of the Fourth-Order Moment $\vm{T}[m]$}
\applab{4th_matrix_moment}

We now evaluate the fourth-order moment $\vm{T}[m] = \E{ \left( \vm{h}[m] \vm{h}^H[m] \right)^2 }$, where we drop the time argument for notational simplicity:
% \begin{IEEEeqnarray}{rCl}
% %  \vm{T} &=& \E{ \vm{h} \vm{h}^H \vm{h} \vm{h}^H }\IEEEnonumber\\
% %  &=& \E{ \left( \bar{\vm{h}} + \tilde{\vm{h}} \right) \left( \bar{\vm{h}} + \tilde{\vm{h}} \right)^H \left( \bar{\vm{h}} + \tilde{\vm{h}} \right) \left( \bar{\vm{h}} + \tilde{\vm{h}} \right)^H }\IEEEnonumber\\
%  \vm{T} &=& \E{ \left( \bar{\vm{h}} \bar{\vm{h}}^H + \tilde{\vm{h}} \tilde{\vm{h}}^H + \bar{\vm{h}} \tilde{\vm{h}}^H + \tilde{\vm{h}} \bar{\vm{h}}^H \right)^2 }\IEEEnonumber\\
%  &\stackrel{(\text{a})}{=}& \E{ \bar{\vm{h}} \bar{\vm{h}}^H \bar{\vm{h}} \bar{\vm{h}}^H } + \E{ \tilde{\vm{h}} \tilde{\vm{h}}^H \tilde{\vm{h}} \tilde{\vm{h}}^H } + \bar{\vm{R}} \tilde{\vm{R}} + \tilde{\vm{R}} \bar{\vm{R}}\IEEEnonumber\\
%  && + \E{ \bar{\vm{h}} \tr{\tilde{\vm{R}}} \bar{\vm{h}}^H } + \E{ \tilde{\vm{h}} \tr{\bar{\vm{R}}} \tilde{\vm{h}}^H }\IEEEnonumber\\
%  &\stackrel{(b)}{=}& \bar{\vm{R}} \tr{\bar{\vm{R}}} + \tilde{\vm{R}}^2 + \tilde{\vm{R}} \tr{\tilde{\vm{R}}} + \bar{\vm{R}} \tilde{\vm{R}}\IEEEnonumber\\
%  && + \tilde{\vm{R}} \bar{\vm{R}} + \bar{\vm{R}} \tr{\tilde{\vm{R}}} + \tilde{\vm{R}} \tr{\bar{\vm{R}}}\IEEEnonumber\\
%  &\stackrel{(c)}{=}& \vm{R} \tr{\vm{R}} + \vm{R}^2 - \bar{\vm{R}}^2 .
%  \eqnlab{app_T}
% \end{IEEEeqnarray}
\begin{IEEEeqnarray}{rCl}
%  \vm{T} &=& \E{ \vm{h} \vm{h}^H \vm{h} \vm{h}^H }\IEEEnonumber\\
%  &=& \E{ \left( \bar{\vm{h}} + \tilde{\vm{h}} \right) \left( \bar{\vm{h}} + \tilde{\vm{h}} \right)^H \left( \bar{\vm{h}} + \tilde{\vm{h}} \right) \left( \bar{\vm{h}} + \tilde{\vm{h}} \right)^H }\IEEEnonumber\\
 \vm{T} &=& \E{ \left( \bar{\vm{h}} \bar{\vm{h}}^H + \tilde{\vm{h}} \tilde{\vm{h}}^H + \bar{\vm{h}} \tilde{\vm{h}}^H + \tilde{\vm{h}} \bar{\vm{h}}^H \right)^2 }\IEEEnonumber\\
 &\stackrel{(\text{a})}{=}& \op{E} \big\{ \bar{\vm{h}} \bar{\vm{h}}^H \bar{\vm{h}} \bar{\vm{h}}^H \big\} + \op{E} \big\{ \tilde{\vm{h}} \tilde{\vm{h}}^H \tilde{\vm{h}} \tilde{\vm{h}}^H \big\} + \bar{\vm{R}} \tilde{\vm{R}} + \tilde{\vm{R}} \bar{\vm{R}} + \op{E} \big\{ \bar{\vm{h}} \op{tr}\big\{\tilde{\vm{R}}\big\} \bar{\vm{h}}^H \big\} + \op{E} \big\{ \tilde{\vm{h}} \op{tr}\big\{\bar{\vm{R}}\big\} \tilde{\vm{h}}^H \big\}\IEEEnonumber\\
 &\stackrel{(b)}{=}& \bar{\vm{R}} \op{tr}\big\{\bar{\vm{R}}\big\} + \tilde{\vm{R}}^2 + \tilde{\vm{R}} \op{tr}\big\{\tilde{\vm{R}}\big\} + \bar{\vm{R}} \tilde{\vm{R}} + \tilde{\vm{R}} \bar{\vm{R}} + \bar{\vm{R}} \op{tr}\big\{\tilde{\vm{R}}\big\} + \tilde{\vm{R}} \op{tr}\big\{\bar{\vm{R}}\big\}\IEEEnonumber\\
 &\stackrel{(c)}{=}& \vm{R} \tr{\vm{R}} + \vm{R}^2 - \bar{\vm{R}}^2 .
 \eqnlab{app_T}
\end{IEEEeqnarray}
In (a), we used $\op{E}\{\bar{\vm{h}}\} = \vm{0}_{N_{\text{TX}} N_{\text{RX}}, 1}$, $\op{E}\{\tilde{\vm{h}}\} = \vm{0}_{N_{\text{TX}} N_{\text{RX}}, 1}$, the mutual independency of $\bar{\vm{h}}$ and $\tilde{\vm{h}}$, and that $\op{E}\{ \tilde{\vm{h}} \tilde{\vm{h}}^T \} = \vm{0}_{N_{\text{TX}} N_{\text{RX}}, N_{\text{TX}} N_{\text{RX}}}$ holds due to properness of $\tilde{\vm{h}}$. In (b), we made use of the fact that $\bar{\vm{h}}^H \bar{\vm{h}} = \op{tr}\{\bar{\vm{R}}\}$, and we used \cite[Th.~1]{Janssen_Gaussian_Matrix_Moments} which yields the following identity for the zero-mean proper Gaussian random vector $\tilde{\vm{h}}$:\looseness=-1
% \begin{IEEEeqnarray}{rCl}
%  \IEEEeqnarraymulticol{3}{l}{ \E{ \tilde{\vm{h}} \tilde{\vm{h}}^H \tilde{\vm{h}} \tilde{\vm{h}}^H } }\IEEEnonumber\\
%  &=& \E{ \tilde{\vm{h}} \tilde{\vm{h}}^H } \E{ \tilde{\vm{h}} \tilde{\vm{h}}^H } + \E{ \tilde{\vm{h}} \E{ \tilde{\vm{h}}^H \tilde{\vm{h}} } \tilde{\vm{h}}^H } .
%  \eqnlab{app_matrix_moment_identity}
% \end{IEEEeqnarray}
\begin{IEEEeqnarray}{rCl}
 \E{ \tilde{\vm{h}} \tilde{\vm{h}}^H \tilde{\vm{h}} \tilde{\vm{h}}^H } &=& \E{ \tilde{\vm{h}} \tilde{\vm{h}}^H } \E{ \tilde{\vm{h}} \tilde{\vm{h}}^H } + \E{ \tilde{\vm{h}} \E{ \tilde{\vm{h}}^H \tilde{\vm{h}} } \tilde{\vm{h}}^H } .
 \eqnlab{app_matrix_moment_identity}
\end{IEEEeqnarray}
In (c), we used $\vm{R} = \bar{\vm{R}} + \tilde{\vm{R}}$.

\section{Sufficient Condition for a Positive Semidefinite $\tilde{\vm{R}}^{(e)}[m]$}
\applab{psd_condition}

In order to derive a sufficient condition for the positive semidefiniteness of $\breve{\vm{R}}_{k}[m], \forall k=1,\ldots,N_{\text{DP}}$ and thus $\tilde{\vm{R}}^{(e)}[m]$, we need to solve the following inequality for $c_k[m], \forall k=1,\ldots,N_{\text{DP}}$ for which we drop the time argument:
\begin{IEEEeqnarray}{rCl}
 \vm{z}^H \left( \breve{\vm{R}}_{k-1} - c_k~ \check{\vm{u}}_k \check{\vm{u}}_k^H\right)\vm{z} &\geq& 0 ,~ \forall \vm{z}\in\mathbb{C}^{N_{\text{TX}} N_{\text{RX}} \times 1} .
 \eqnlab{app_c_cond_1}
\end{IEEEeqnarray}
The case $\vm{z} = \vm{0}_{N_{\text{TX}} N_{\text{RX}}, 1}$ is trivially satisfied. In case $\vm{z} \neq \vm{0}_{N_{\text{TX}} N_{\text{RX}}, 1}$, we first consider non-singular $\breve{\vm{R}}_{k-1}$. We define $\check{\vm{z}} = \breve{\vm{R}}_{k-1}^{\frac{1}{2}} \vm{z}$ and rearrange \eqnref{app_c_cond_1} to obtain
\begin{IEEEeqnarray}{rCl}
 \frac{c_k~ \vm{z}^H \check{\vm{u}}_k \check{\vm{u}}_k^H \vm{z}}{\vm{z}^H \breve{\vm{R}}_{k-1} \vm{z}} &=& \frac{c_k~ \check{\vm{z}}^H \breve{\vm{R}}_{k-1}^{-\frac{1}{2}} \check{\vm{u}}_k \check{\vm{u}}_k^H \breve{\vm{R}}_{k-1}^{-\frac{1}{2}} \check{\vm{z}}}{\check{\vm{z}}^H \check{\vm{z}}} \leq 1 .
 \eqnlab{app_c_cond_2}
\end{IEEEeqnarray}
The matrix $\breve{\vm{R}}_{k-1}^{-\frac{1}{2}} \check{\vm{u}}_k \check{\vm{u}}_k^H \breve{\vm{R}}_{k-1}^{-\frac{1}{2}}$ is positive semidefinite with rank one such that, with the Rayleigh-Ritz theorem \cite[Th.~4.2.2]{Horn_Matrix_Analysis}, we have
% \begin{IEEEeqnarray}{rCl}
%  0 &\leq& \frac{\check{\vm{z}}^H \breve{\vm{R}}_{k-1}^{-\frac{1}{2}} \check{\vm{u}}_k \check{\vm{u}}_k^H \breve{\vm{R}}_{k-1}^{-\frac{1}{2}}\check{\vm{z}}}{\check{\vm{z}}^H \check{\vm{z}}}\IEEEnonumber\\
%  &\leq& \lambda_{\text{max}} \left( \breve{\vm{R}}_{k-1}^{-\frac{1}{2}} \check{\vm{u}}_k \check{\vm{u}}_k^H \breve{\vm{R}}_{k-1}^{-\frac{1}{2}}\right) .
%  \eqnlab{app_rayleigh_ritz}
% \end{IEEEeqnarray}
\begin{IEEEeqnarray}{rCl}
 0 &\leq& \frac{\check{\vm{z}}^H \breve{\vm{R}}_{k-1}^{-\frac{1}{2}} \check{\vm{u}}_k \check{\vm{u}}_k^H \breve{\vm{R}}_{k-1}^{-\frac{1}{2}}\check{\vm{z}}}{\check{\vm{z}}^H \check{\vm{z}}} \leq \lambda_{\text{max}} \left( \breve{\vm{R}}_{k-1}^{-\frac{1}{2}} \check{\vm{u}}_k \check{\vm{u}}_k^H \breve{\vm{R}}_{k-1}^{-\frac{1}{2}}\right) .
 \eqnlab{app_rayleigh_ritz}
\end{IEEEeqnarray}
Finally, with \eqnref{app_c_cond_2} and \eqnref{app_rayleigh_ritz}, we obtain
% \begin{IEEEeqnarray}{rCl}
%  c_k &\leq& \lambda_{\text{max}}^{-1} \! \left( \! \breve{\vm{R}}_{k-1}^{-\frac{1}{2}} \check{\vm{u}}_k \check{\vm{u}}_k^H \breve{\vm{R}}_{k-1}^{-\frac{1}{2}} \! \right)\IEEEnonumber\\
%  &=& \left( \! \check{\vm{u}}_k^H \breve{\vm{R}}_{k-1}^{-1} \check{\vm{u}}_k \! \right)^{-1} ,~ \forall k=1,\ldots,N_{\text{DP}}
%  \eqnlab{app_c_cond_3}
% \end{IEEEeqnarray}
\begin{IEEEeqnarray}{rCl}
 c_k &\leq& \lambda_{\text{max}}^{-1} \left( \breve{\vm{R}}_{k-1}^{-\frac{1}{2}} \check{\vm{u}}_k \check{\vm{u}}_k^H \breve{\vm{R}}_{k-1}^{-\frac{1}{2}} \right) = \left( \check{\vm{u}}_k^H \breve{\vm{R}}_{k-1}^{-1} \check{\vm{u}}_k \right)^{-1} ,~ \forall k=1,\ldots,N_{\text{DP}}
 \eqnlab{app_c_cond_3}
\end{IEEEeqnarray}
which is a necessary and sufficient condition for $\breve{\vm{R}}_{k}, \forall k=1,\ldots,N_{\text{DP}}$ to be positive semidefinite if $\breve{\vm{R}}_{k-1}, \forall k=1,\ldots,N_{\text{DP}}$ is non-singular. In the case of a singular $\breve{\vm{R}}_{k-1}$, we set $c_k = 0$. We thus obtain a sufficient condition for $\tilde{\vm{R}}^{(e)}$ to be positive semidefinite. We note that \eqnref{app_c_cond_3} (for non-singular $\breve{\vm{R}}_{k-1}, \forall k=1,\ldots,N_{\text{DP}}$) can also be derived based on \cite[Th.~7.7.7]{Horn_Matrix_Analysis}.

\section{Approximate Evaluation of the MI}
\applab{MI_approx}

The approximate evaluation of the MI relies on a multivariate Taylor series expansion. We consider a complex function $f(\vm{a},\vm{a}^{\ast})$ with complex column vector arguments $\vm{a}$ and $\vm{a}^{\ast}$ of lengths $N^2$. We note that $\vm{a}^{\ast}$ is the complex conjugate of $\vm{a}$. The second-order approximation of $\vm{a}$ and $\vm{a}^{\ast}$ at $\vm{a}_0$ and $\vm{a}^{\ast}_0$, respectively, is given by \cite{Hjorungnes_Complex_Matrix_Deriv_App}
% \begin{IEEEeqnarray}{rCl}
%  f(\vm{a},\vm{a}^{\ast}) &\approx& f(\vm{a}_0,\vm{a}^{\ast}_0) + \left.\frac{\partial f}{\partial \vm{a}}\right|_{\vm{a}=\vm{a}_0,\vm{a}^{\ast}=\vm{a}^{\ast}_0} \cdot (\vm{a}-\vm{a}_0)\IEEEnonumber\\
%  && + \left.\frac{\partial f}{\partial \vm{a}^{\ast}}\right|_{\vm{a}=\vm{a}_0,\vm{a}^{\ast}=\vm{a}^{\ast}_0} \cdot (\vm{a}^{\ast}-\vm{a}_0^{\ast})\IEEEnonumber\\
%  && + \frac{1}{2} (\vm{a}-\vm{a}_0)^H \cdot \vm{H}_f^{cs}(\vm{a}_0,\vm{a}^{\ast}_0) \cdot (\vm{a}-\vm{a}_0)\IEEEnonumber\\
%  && + \frac{1}{2} (\vm{a}-\vm{a}_0)^T \cdot {\vm{H}_f^{cs}}^T(\vm{a}_0,\vm{a}^{\ast}_0) \cdot (\vm{a}^{\ast}-\vm{a}_0^{\ast})\IEEEnonumber\\
%  && + \frac{1}{2} (\vm{a}-\vm{a}_0)^H \cdot \vm{H}_f^{cc}(\vm{a}_0,\vm{a}^{\ast}_0) \cdot (\vm{a}^{\ast}-\vm{a}_0^{\ast})\IEEEnonumber\\
%  && + \frac{1}{2} (\vm{a}-\vm{a}_0)^T \cdot \vm{H}_f^{ss}(\vm{a}_0,\vm{a}^{\ast}_0) \cdot (\vm{a}-\vm{a}_0)
%  \eqnlab{app_second_order_approx_1}
% \end{IEEEeqnarray}
\begin{IEEEeqnarray}{rCl}
 f(\vm{a},\vm{a}^{\ast}) &\approx& f(\vm{a}_0,\vm{a}^{\ast}_0) + \left.\frac{\partial f}{\partial \vm{a}}\right|_{\vm{a}=\vm{a}_0,\vm{a}^{\ast}=\vm{a}^{\ast}_0} \cdot (\vm{a}-\vm{a}_0) + \left.\frac{\partial f}{\partial \vm{a}^{\ast}}\right|_{\vm{a}=\vm{a}_0,\vm{a}^{\ast}=\vm{a}^{\ast}_0} \cdot (\vm{a}^{\ast}-\vm{a}_0^{\ast})\IEEEnonumber\\
 && + \frac{1}{2} (\vm{a}-\vm{a}_0)^H \cdot \vm{H}_f^{cs}(\vm{a}_0,\vm{a}^{\ast}_0) \cdot (\vm{a}-\vm{a}_0) + \frac{1}{2} (\vm{a}-\vm{a}_0)^T \cdot {\vm{H}_f^{sc}}(\vm{a}_0,\vm{a}^{\ast}_0) \cdot (\vm{a}^{\ast}-\vm{a}_0^{\ast})\IEEEnonumber\\
 && + \frac{1}{2} (\vm{a}-\vm{a}_0)^H \cdot \vm{H}_f^{cc}(\vm{a}_0,\vm{a}^{\ast}_0) \cdot (\vm{a}^{\ast}-\vm{a}_0^{\ast}) + \frac{1}{2} (\vm{a}-\vm{a}_0)^T \cdot \vm{H}_f^{ss}(\vm{a}_0,\vm{a}^{\ast}_0) \cdot (\vm{a}-\vm{a}_0)\IEEEeqnarraynumspace
 \eqnlab{app_second_order_approx_1}
\end{IEEEeqnarray}
with the row vector $\partial f / \partial \vm{a}$ defined by $[\partial f / \partial \vm{a}]_{1,k} = \partial f / \partial [\vm{a}]_{k,1}$ for $k=1,\ldots,N$ and the $N^2 \times N^2$ Hessian matrices
% \begin{IEEEeqnarray}{rCl}
%  \vm{H}_f^{cs}(\vm{a}_0,\vm{a}^{\ast}_0) &=& \left. \frac{\partial}{\partial \vm{a}} \left( \frac{\partial f}{\partial \vm{a}^{\ast}} \right)^T \right|_{\vm{a}=\vm{a}_0,\vm{a}^{\ast}=\vm{a}^{\ast}_0}\\
%  \vm{H}_f^{sc}(\vm{a}_0,\vm{a}^{\ast}_0) &=& \left( \vm{H}_f^{cs}(\vm{a}_0,\vm{a}^{\ast}_0) \right)^T\\
%  \vm{H}_f^{cc}(\vm{a}_0,\vm{a}^{\ast}_0) &=& \left. \frac{\partial}{\partial \vm{a}^{\ast}} \left( \frac{\partial f}{\partial \vm{a}^{\ast}} \right)^T \right|_{\vm{a}=\vm{a}_0,\vm{a}^{\ast}=\vm{a}^{\ast}_0}\\
%  \vm{H}_f^{ss}(\vm{a}_0,\vm{a}^{\ast}_0) &=& \left. \frac{\partial}{\partial \vm{a}} \left( \frac{\partial f}{\partial \vm{a}} \right)^T \right|_{\vm{a}=\vm{a}_0,\vm{a}^{\ast}=\vm{a}^{\ast}_0} .
%  \eqnlab{app_hessians}
% \end{IEEEeqnarray}
\begin{IEEEeqnarray}{rCl}
 \vm{H}_f^{cs}(\vm{a}_0,\vm{a}^{\ast}_0) = \left. \frac{\partial}{\partial \vm{a}} \left( \frac{\partial f}{\partial \vm{a}^{\ast}} \right)^T \right|_{\vm{a}=\vm{a}_0,\vm{a}^{\ast}=\vm{a}^{\ast}_0} &;\hspace{1cm}&
 \vm{H}_f^{sc}(\vm{a}_0,\vm{a}^{\ast}_0) = \left( \vm{H}_f^{cs}(\vm{a}_0,\vm{a}^{\ast}_0) \right)^T\IEEEnonumber\\
 \vm{H}_f^{cc}(\vm{a}_0,\vm{a}^{\ast}_0) = \left. \frac{\partial}{\partial \vm{a}^{\ast}} \left( \frac{\partial f}{\partial \vm{a}^{\ast}} \right)^T \right|_{\vm{a}=\vm{a}_0,\vm{a}^{\ast}=\vm{a}^{\ast}_0} &;\hspace{1cm}&
 \vm{H}_f^{ss}(\vm{a}_0,\vm{a}^{\ast}_0) = \left. \frac{\partial}{\partial \vm{a}} \left( \frac{\partial f}{\partial \vm{a}} \right)^T \right|_{\vm{a}=\vm{a}_0,\vm{a}^{\ast}=\vm{a}^{\ast}_0} .
 \eqnlab{app_hessians}
\end{IEEEeqnarray}
We now consider the function $f(\vm{a},\vm{a}^{\ast}) = f(\vm{a}) = \ln \det{\vm{A}}$ with $\vm{a} = \vect{\vm{A}}$ and the $N \times N$ matrix $\vm{A}$. By using $f(\vm{a})$ in \eqnref{app_second_order_approx_1} with $\vm{a}_0 = \E{\vm{a}}$ and applying the expectation operator, we obtain the second-order approximation
\begin{IEEEeqnarray}{rCl}
 \E{f(\vm{a})} &\approx& f(\E{\vm{a}}) + \frac{1}{2} \tr{ \E{(\vm{a}-\E{\vm{a}}) (\vm{a}-\E{\vm{a}})^T} \vm{H}_f^{ss}(\E{\vm{a}}) }
 \eqnlab{app_second_order_approx_2}
\end{IEEEeqnarray}
where we used that only the first two and the last term in \eqnref{app_second_order_approx_1} are non-zero. The Hessian matrix $\vm{H}_f^{ss}(\E{\vm{a}},\E{\vm{a}^{\ast}}) = \vm{H}_f^{ss}(\E{\vm{a}})$ is given by \cite{Hjorungnes_Complex_Matrix_Diff_Results}
\begin{IEEEeqnarray}{rCl}
 \vm{H}_f^{ss}(\E{\vm{a}}) &=& - \vm{K}_{N,N} \left( \left( \E{\vm{A}} \right)^{-T} \otimes \left( \E{\vm{A}} \right)^{-1} \right) .
 \eqnlab{app_hessian_ss}
\end{IEEEeqnarray}
For $\vm{A} = \vm{B} + \vm{C} \vm{D} \vm{E}$ with deterministic $N \times N$ matrices $\vm{B}$, $\vm{C}$, and $\vm{E}$, \eqnref{app_second_order_approx_2} can be written as
% \begin{IEEEeqnarray}{rCl}
%  \E{f(\vm{a})} &\approx& f(\E{\vm{a}}) + \frac{1}{2} \op{tr} \big\{ \left( \vm{E}^T \otimes \vm{C} \right)\IEEEnonumber\\
%  && \times \E{(\vm{d}-\E{\vm{d}}) (\vm{d}-\E{\vm{d}})^T}\IEEEnonumber\\
%  && \times \left( \vm{E} \otimes \vm{C}^T \right) \vm{H}_f^{ss}(\E{\vm{a}}) \big\}
%  \eqnlab{app_second_order_approx_3}
% \end{IEEEeqnarray}
\begin{IEEEeqnarray}{rCl}
 \E{f(\vm{a})} &\stackrel{(\text{a})}{\approx}& f(\E{\vm{a}}) + \frac{1}{2} \op{tr} \big\{ \left( \vm{E}^T \otimes \vm{C} \right) \E{(\vect{\vm{D}-\E{\vm{D}}}) (\vect{\vm{D}-\E{\vm{D}}})^T}\IEEEnonumber\\
 &&\times \left( \vm{E} \otimes \vm{C}^T \right) \vm{H}_f^{ss}(\E{\vm{a}}) \big\}\IEEEnonumber\\
 &\stackrel{(\text{b})}{=}& f(\E{\vm{a}}) - \frac{1}{2} \op{tr} \big\{ \left( \vm{E}^T \otimes \vm{C} \right) \E{(\vect{\vm{D}-\E{\vm{D}}}) (\vect{\vm{D}-\E{\vm{D}}})^T} \vm{K}_{N,N}\IEEEnonumber\\
 &&\times \left( \vm{C}^T \otimes \vm{E} \right) \left( \left( \E{\vm{A}} \right)^{-T} \otimes \left( \E{\vm{A}} \right)^{-1} \right) \big\}\IEEEnonumber\\
%  &=& f(\E{\vm{a}}) - \frac{1}{2} \op{tr} \big\{ \E{(\vect{\vm{D}-\E{\vm{D}}}) (\vect{\vm{D}^H-\E{\vm{D}^H}})^H}\IEEEnonumber\\
%  &&\times \left( \vm{C}^T \otimes \vm{E} \right) \left( \left( \E{\vm{A}} \right)^{-T} \otimes \left( \E{\vm{A}} \right)^{-1} \right) \left( \vm{E}^T \otimes \vm{C} \right) \big\}\IEEEnonumber\\
 &\stackrel{(\text{c})}{=}& f(\E{\vm{a}}) - \frac{1}{2} \op{tr} \big\{ \E{(\vect{\vm{D}-\E{\vm{D}}}) (\vect{\vm{D}^H-\E{\vm{D}^H}})^H}\IEEEnonumber\\
 &&\times \left( \vm{E} \left( \E{\vm{A}} \right)^{-1} \vm{C} \right)^T \otimes \left( \vm{E} \left( \E{\vm{A}} \right)^{-1} \vm{C} \right) \big\} .
 \eqnlab{app_second_order_approx_3}
\end{IEEEeqnarray}
In (a), we applied $\vect{\vm{C} \vm{D} \vm{E}} = \left( \vm{E}^T \otimes \vm{C} \right) \vect{\vm{D}}$ \cite[Lemma 4.3.1]{Horn_Matrix_Analysis_2}. 
% \begin{IEEEeqnarray}{rCl}
%  \vect{\vm{C} \vm{D} \vm{E}} &=& \left( \vm{E}^T \otimes \vm{C} \right) \vect{\vm{D}} .
%  \eqnlab{app_vec_trick}
% \end{IEEEeqnarray}
In (b), we inserted \eqnref{app_hessian_ss} and used $\left( \vm{E} \otimes \vm{C}^T \right) \vm{K}_{N,N} = \vm{K}_{N,N} \left( \vm{C}^T \otimes \vm{E} \right)$ \cite[Th.~3.1~(viii)]{Magnus_Commutation_Matrix}.
% \begin{IEEEeqnarray}{rCl}
%  \left( \vm{E} \otimes \vm{C}^T \right) \vm{K}_{N,N} &=& \vm{K}_{N,N} \left( \vm{C}^T \otimes \vm{E} \right) .
%  \eqnlab{app_comm_trick_8}
% \end{IEEEeqnarray}
Finally, in (c), we used $\vm{K}_{N,N}^T = \vm{K}_{N,N}$ \cite[Th.~3.1~(ii)]{Magnus_Commutation_Matrix} and 
$\left( \vm{A} \otimes \vm{B} \right) \left( \vm{C} \otimes \vm{D} \right) = \left( \vm{A} \vm{C} \right) \otimes \left( \vm{B} \vm{D} \right)$ \cite[Lemma~4.2.10]{Horn_Matrix_Analysis_2}.
% \cite[Lemma~4.2.10]{Horn_Matrix_Analysis_2}
% \begin{IEEEeqnarray}{rCl}
%  \left( \vm{A} \otimes \vm{B} \right) \left( \vm{C} \otimes \vm{D} \right) &=& \left( \vm{A} \vm{C} \right) \otimes \left( \vm{B} \vm{D} \right) .
%  \eqnlab{app_kron_trick}
% \end{IEEEeqnarray}

\section{Approximate Evaluation of the MI by Means of the Proposed Channel Model}
\applab{MI_approx_cm}

We restate \eqnref{MI_approx_Z} as a function of the parameters obtained in the channel decomposition in \secref{cm_decomposition}\ie $\bar{\vm{R}}[m]$ and $\tilde{\vm{R}}[m]$, only. To simplify notation, we drop the time argument for the remainder of this appendix. First, we rewrite \eqnref{MI_approx_Z}:
\begin{IEEEeqnarray}{rCl}
 \vm{Z} &\stackrel{(\text{a})}{=}& \E{ \vect{ \bar{\vm{H}}^H \bar{\vm{H}} + \tilde{\vm{H}}^H \tilde{\vm{H}} + \bar{\vm{H}}^H \tilde{\vm{H}} + \tilde{\vm{H}}^H \bar{\vm{H}} } \left( \vect{ \bar{\vm{H}}^H \bar{\vm{H}} + \tilde{\vm{H}}^H \tilde{\vm{H}} + \bar{\vm{H}}^H \tilde{\vm{H}} + \tilde{\vm{H}}^H \bar{\vm{H}} } \right)^H }\IEEEnonumber\\
 && - \vect{ \bar{\vm{R}}_{\text{TX}}^{\ast} + \tilde{\vm{R}}_{\text{TX}}^{\ast} } \left( \vect{ \bar{\vm{R}}_{\text{TX}}^{\ast} + \tilde{\vm{R}}_{\text{TX}}^{\ast} } \right)^H\IEEEnonumber\\
 &\stackrel{(\text{b})}{=}& \E{ \vect{\bar{\vm{H}}^H \bar{\vm{H}}} \left( \vect{\bar{\vm{H}}^H \bar{\vm{H}}} \right)^H } - \vect{\bar{\vm{R}}_{\text{TX}}^{\ast}} \left( \vect{\bar{\vm{R}}_{\text{TX}}^{\ast}} \right)^H\IEEEnonumber\\
 && + \E{ \vect{\tilde{\vm{H}}^H \tilde{\vm{H}}} \left( \vect{\tilde{\vm{H}}^H \tilde{\vm{H}}} \right)^H } - \vect{\tilde{\vm{R}}_{\text{TX}}^{\ast}} \left( \vect{\tilde{\vm{R}}_{\text{TX}}^{\ast}} \right)^H\IEEEnonumber\\
 && + \E{ \vect{\bar{\vm{H}}^H \tilde{\vm{H}}} \left( \vect{\bar{\vm{H}}^H \tilde{\vm{H}}} \right)^H } + \E{ \vect{\tilde{\vm{H}}^H \bar{\vm{H}}} \left( \vect{\tilde{\vm{H}}^H \bar{\vm{H}}} \right)^H }
 \eqnlab{app_MI_approx_param_rewrite_1}
\end{IEEEeqnarray}
with $\bar{\vm{R}}_{\text{TX}} = \op{E}\{ \bar{\vm{H}}^T \bar{\vm{H}}^{\ast} \}$ and $\tilde{\vm{R}}_{\text{TX}} = \op{E}\{ \tilde{\vm{H}}^T \tilde{\vm{H}}^{\ast} \}$. In (a), we applied $\vm{H} = \bar{\vm{H}} + \tilde{\vm{H}}$ and $\vm{R}_{\text{TX}} = \bar{\vm{R}}_{\text{TX}} + \tilde{\vm{R}}_{\text{TX}}$. In (b), we used the properness of $\tilde{\vm{H}}$ to establish
\begin{IEEEeqnarray}{rCl}
 \E{\vect{\bar{\vm{H}}^H \tilde{\vm{H}}} \left( \vect{\tilde{\vm{H}}^H \bar{\vm{H}}} \right)^H} &=& \vm{0}_{N_{\text{TX}}^2,N_{\text{TX}}^2} .
 \eqnlab{app_MI_approx_param_rewrite_2}
\end{IEEEeqnarray}
We now have
\begin{IEEEeqnarray}{rCl}
 \E{\vect{\tilde{\vm{H}}^H \bar{\vm{H}}} \left( \vect{\tilde{\vm{H}}^H \bar{\vm{H}}} \right)^H} &\stackrel{(\text{a})}{=}& \E{ \left( \vm{I}_{N_{\text{TX}}} \otimes \tilde{\vm{H}}^H \right) \vect{\bar{\vm{H}}} \left( \vect{\bar{\vm{H}}}\right)^H \left( \vm{I}_{N_{\text{TX}}} \otimes \tilde{\vm{H}} \right) }\IEEEnonumber\\
 &=& \E{ \left( \vm{I}_{N_{\text{TX}}} \otimes \tilde{\vm{H}}^H \right) \bar{\vm{R}} \left( \vm{I}_{N_{\text{TX}}} \otimes \tilde{\vm{H}} \right) }
 \eqnlab{app_MI_approx_param_rewrite_3}
\end{IEEEeqnarray}
where, in (a), we used \cite[Lemma~4.3.1]{Horn_Matrix_Analysis_2}. % as in \eqnref{app_vec_trick}. 
Similarly, we have
\begin{IEEEeqnarray}{rCl}
 \IEEEeqnarraymulticol{3}{l}{ \E{\vect{\bar{\vm{H}}^H \tilde{\vm{H}}} \left( \vect{\bar{\vm{H}}^H \tilde{\vm{H}}} \right)^H} }\IEEEnonumber\\
%  &=& \E{ \left( \tilde{\vm{H}}^T \otimes \vm{I}_{N_{\text{TX}}} \right) \vect{\bar{\vm{H}}^H} \left( \vect{\bar{\vm{H}}^H}\right)^H \left( \tilde{\vm{H}}^{\ast} \otimes \vm{I}_{N_{\text{TX}}} \right) }\IEEEnonumber\\
 &\stackrel{(\text{a})}{=}& \left( \E{ \left( \tilde{\vm{H}}^H \otimes \vm{I}_{N_{\text{TX}}} \right) \vm{K}_{N_{\text{TX}},N_{\text{TX}}} \vect{\bar{\vm{H}}} \left( \vect{\bar{\vm{H}}}\right)^H \vm{K}_{N_{\text{TX}},N_{\text{TX}}} \left( \tilde{\vm{H}} \otimes \vm{I}_{N_{\text{TX}}} \right) } \right)^{\ast}\IEEEnonumber\\
 &\stackrel{(\text{b})}{=}& \vm{K}_{N_{\text{TX}},N_{\text{TX}}} \left( \E{ \left( \vm{I}_{N_{\text{TX}}} \otimes \tilde{\vm{H}}^H \right) \bar{\vm{R}} \left( \vm{I}_{N_{\text{TX}}} \otimes \tilde{\vm{H}} \right) } \right)^{\ast} \vm{K}_{N_{\text{TX}},N_{\text{TX}}}
 \eqnlab{app_MI_approx_param_rewrite_4}
\end{IEEEeqnarray}
where, in (a), we used \cite[Th.~3.1~(ii)]{Magnus_Commutation_Matrix}, and, in (b), we used \cite[Th.~3.1~(viii)]{Magnus_Commutation_Matrix}. Next, we have
\begin{IEEEeqnarray}{rCl}
 \IEEEeqnarraymulticol{3}{l}{ \E{ \vect{\tilde{\vm{H}}^H \tilde{\vm{H}}} \left( \vect{\tilde{\vm{H}}^H \tilde{\vm{H}}} \right)^H } }\IEEEnonumber\\
%  &=& \E{ \left( \vm{I}_{N_{\text{TX}}} \otimes \tilde{\vm{H}}^H \right) \vect{\tilde{\vm{H}}} \left( \vect{\tilde{\vm{H}}}\right)^H \left( \vm{I}_{N_{\text{TX}}} \otimes \tilde{\vm{H}}^H \right)^H }\IEEEnonumber\\
 &\stackrel{(\text{a})}{=}& \E{\left( \vm{I}_{N_{\text{TX}}} \otimes \tilde{\vm{H}}^H \right) \vect{\tilde{\vm{H}}} } \E{ \left( \vect{\tilde{\vm{H}}} \right)^H \left( \vm{I}_{N_{\text{TX}}} \otimes \tilde{\vm{H}}^H \right)^H }\IEEEnonumber\\
 && + \E{ \left( \vm{I}_{N_{\text{TX}}} \otimes \tilde{\vm{H}}^H \right) \tilde{\vm{R}} \left(  \vm{I}_{N_{\text{TX}}} \otimes \tilde{\vm{H}} \right) }\IEEEnonumber\\
%  &=& \E{ \vect{\tilde{\vm{H}}^H \tilde{\vm{H}}} } \E{ \left( \vect{\tilde{\vm{H}}^H \tilde{\vm{H}}} \right)^H } + \E{ \left( \vm{I}_{N_{\text{TX}}} \otimes \tilde{\vm{H}}^H \right) \tilde{\vm{R}} \left( \vm{I}_{N_{\text{TX}}} \otimes \tilde{\vm{H}} \right) }\IEEEnonumber\\
 &=& \vect{\tilde{\vm{R}}_{\text{TX}}^{\ast}} \left( \vect{\tilde{\vm{R}}_{\text{TX}}^{\ast}} \right)^H + \E{ \left( \vm{I}_{N_{\text{TX}}} \otimes \tilde{\vm{H}}^H \right) \tilde{\vm{R}} \left( \vm{I}_{N_{\text{TX}}} \otimes \tilde{\vm{H}} \right) }
 \eqnlab{app_MI_approx_param_rewrite_5}
\end{IEEEeqnarray}
where, in (a), we used \cite[Th.~1]{Janssen_Gaussian_Matrix_Moments} with the properness of $\tilde{\vm{H}}$. In order to evaluate \eqnref{app_MI_approx_param_rewrite_3}, \eqnref{app_MI_approx_param_rewrite_4}, and \eqnref{app_MI_approx_param_rewrite_5}, we use that
\begin{IEEEeqnarray}{rCl}
 \vect{ \E{ \left( \vm{I}_{N_{\text{TX}}} \otimes \tilde{\vm{H}}^H \right) \vm{A} \left( \vm{I}_{N_{\text{TX}}} \otimes \tilde{\vm{H}} \right) } } &=& \left( \vm{I}_{N_{\text{TX}}} \otimes \E{ \tilde{\vm{H}}^T \otimes \vm{I}_{N_{\text{TX}}} \otimes \tilde{\vm{H}}^H } \right) \vect{\vm{A}}\IEEEnonumber\\
 &=& \left( \vm{I}_{N_{\text{TX}}} \otimes \vm{Y} \right) \vect{\vm{A}}
 \eqnlab{app_MI_approx_param_rewrite_6}
\end{IEEEeqnarray}
holds for a deterministic $N_{\text{TX}} N_{\text{RX}} \times N_{\text{TX}} N_{\text{RX}}$ matrix $\vm{A}$. Here, the $N_{\text{TX}}^3 \times N_{\text{TX}} N_{\text{RX}}^2$ block matrix $\vm{Y}$ contains $\vm{I}_{N_{\text{TX}}} \otimes \vm{X}_{k,l}$ in the $k$th row-partition and the $l$th column-partition for $k = 1, \ldots, N_{\text{TX}}$ and $l = 1, \ldots, N_{\text{RX}}$. The $N_{\text{TX}} \times N_{\text{RX}}$ matrix $\vm{X}_{k,l}$ is defined by $\big[ \vm{X}_{k,l} \big]_{p,q} = \big[ \tilde{\vm{R}} \big]_{(k-1)N_{\text{RX}}+l, (p-1)N_{\text{RX}}+q}$ for $p = 1, \ldots, N_{\text{TX}}$ and $q = 1, \ldots, N_{\text{RX}}$. For the DP case where only the co-polarized sub-links can be affected by dominant components, we can write $\bar{\vm{H}} = \bar{\vm{H}}_1 + \bar{\vm{H}}_2$ with
% \begin{IEEEeqnarray}{rCl}
%  \bar{\vm{H}} &=& \bar{\vm{H}}_1 + \bar{\vm{H}}_2
% \end{IEEEeqnarray}
% with
\begin{IEEEeqnarray}{rCl}
 \bar{\vm{H}}_1 = \begin{bmatrix} \bar{\vm{H}}_{\text{VV}} & \vm{0}_{\frac{N_{\text{RX}}}{2}, \frac{N_{\text{TX}}}{2}}\\
 \vm{0}_{\frac{N_{\text{RX}}}{2}, \frac{N_{\text{TX}}}{2}} & \vm{0}_{\frac{N_{\text{RX}}}{2}, \frac{N_{\text{TX}}}{2}}
 \end{bmatrix} &;\hspace{1cm}&
 \bar{\vm{H}}_2 = \begin{bmatrix}
 \vm{0}_{\frac{N_{\text{RX}}}{2}, \frac{N_{\text{TX}}}{2}} & \vm{0}_{\frac{N_{\text{RX}}}{2}, \frac{N_{\text{TX}}}{2}}\\
 \vm{0}_{\frac{N_{\text{RX}}}{2}, \frac{N_{\text{TX}}}{2}} & \bar{\vm{H}}_{\text{HH}}
 \end{bmatrix} .
\end{IEEEeqnarray}
Obviously, we have $\bar{\vm{H}}_1^H \bar{\vm{H}}_2 = \vm{0}_{N_{\text{TX}}, N_{\text{TX}}}$ and $\bar{\vm{H}}^H \bar{\vm{H}} = \bar{\vm{H}}_1^H \bar{\vm{H}}_1 + \bar{\vm{H}}_2^H \bar{\vm{H}}_2$. Furthermore, with  \eqnref{channel_model_dominant_rewrite}, we have $\bar{\vm{R}}_{\text{TX}}^{\ast} = \E{\bar{\vm{H}}_1^H \bar{\vm{H}}_1 + \bar{\vm{H}}_2^H \bar{\vm{H}}_2} = \bar{\vm{H}}^H \bar{\vm{H}}$. It thus follows that
\begin{IEEEeqnarray}{rCl}
 \E{ \vect{\bar{\vm{H}}^H \bar{\vm{H}}} \left( \vect{\bar{\vm{H}}^H \bar{\vm{H}}} \right)^H } &=& \vect{\bar{\vm{R}}_{\text{TX}}^{\ast}} \left( \vect{\bar{\vm{R}}_{\text{TX}}^{\ast}} \right)^H .
 \eqnlab{app_MI_approx_param_rewrite_7}
\end{IEEEeqnarray}
Clearly, the same result holds in the SP case. At last, using \eqnref{app_MI_approx_param_rewrite_1} with \eqnref{app_MI_approx_param_rewrite_3}, \eqnref{app_MI_approx_param_rewrite_4}, \eqnref{app_MI_approx_param_rewrite_5}, \eqnref{app_MI_approx_param_rewrite_6}, and \eqnref{app_MI_approx_param_rewrite_7}, we obtain the result in \eqnref{MI_approx_Z_2}.
% \begin{IEEEeqnarray}{rCl}
%  \vm{Z} &=& \check{\vm{Z}} + \vm{K}_{N_{\text{TX}},N_{\text{TX}}} \bar{\vm{Z}}^{\ast} \vm{K}_{N_{\text{TX}},N_{\text{TX}}}
%  \eqnlab{app_MI_approx_param_rewrite_8}
% \end{IEEEeqnarray}
% with
% \begin{IEEEeqnarray}{rCl}
%  \vect{\check{\vm{Z}}} &=& \left( \vm{I}_{N_{\text{TX}}} \otimes \vm{Y} \right) \vect{\vm{R}}\\
%  \vect{\bar{\vm{Z}}} &=& \left( \vm{I}_{N_{\text{TX}}} \otimes \vm{Y} \right) \vect{\bar{\vm{R}}} .
% \end{IEEEeqnarray}

\section{Lower Bound on the Approximate Evaluation of the MI}
\applab{MI_approx_LB}

In order to lower-bound the approximate MI \eqnref{MI_approx}, we find an upper bound for the trace in the second term of \eqnref{MI_approx} for the case that the eigenvectors of $\vm{R}^{\ast}_{\text{TX}}[m]$ form the precoding for the $N_{\text{st}}$ transmitted streams. We drop the time argument in the following derivation. Using the eigendecompositions $\vm{R}_{\text{TX}}^{\ast} = \vm{U}_{\text{TX}} \vm{\Lambda}_{\text{TX}} \vm{U}_{\text{TX}}^H$ and $\vm{Q} = \vm{U}_{\text{TX}} \vm{\Lambda}_{Q} \vm{U}_{\text{TX}}^H$, we can write
\begin{IEEEeqnarray}{rCl}
 \IEEEeqnarraymulticol{3}{l}{ \tr{ \vm{Z} \left( \left( \vm{Q} \left( \vm{I}_{N_{\text{TX}}} + \rho \vm{R}_{\text{TX}}^{\ast} \vm{Q} \right)^{-1} \right)^T \otimes \left( \vm{Q} \left( \vm{I}_{N_{\text{TX}}} + \rho \vm{R}_{\text{TX}}^{\ast} \vm{Q} \right)^{-1} \right) \right) } }\IEEEnonumber\\
%  &=& \tr{ \vm{Z} \left( \left( \vm{U}_{\text{TX}} \vm{\Lambda}_{Q} \left( \vm{I}_{N_{\text{TX}}} + \rho \vm{\Lambda}_{\text{TX}} \vm{\Lambda}_{Q} \right)^{-1} \vm{U}_{\text{TX}}^H \right)^T \otimes \left( \vm{U}_{\text{TX}} \vm{\Lambda}_{Q} \left( \vm{I}_{N_{\text{TX}}} + \rho \vm{\Lambda}_{\text{TX}} \vm{\Lambda}_{Q} \right)^{-1} \vm{U}_{\text{TX}}^H \right) \right) }\IEEEnonumber\\
 &\stackrel{(\text{a})}{=}& \tr{ \vm{Z} ( \vm{U}_{\text{TX}}^{\ast} \otimes \vm{U}_{\text{TX}} ) \! \left( \! \left( \vm{\Lambda}_{Q} \left( \vm{I}_{N_{\text{TX}}} + \rho \vm{\Lambda}_{\text{TX}} \vm{\Lambda}_{Q} \right)^{-1} \right)^T \!\! \otimes \! \left( \vm{\Lambda}_{Q} \left( \vm{I}_{N_{\text{TX}}} + \rho \vm{\Lambda}_{\text{TX}} \vm{\Lambda}_{Q} \right)^{-1} \right) \! \right) \! ( \vm{U}_{\text{TX}}^T \otimes \vm{U}_{\text{TX}}^H ) }\IEEEnonumber\\
 &\stackrel{(\text{b})}{=}& \op{tr} \big\{ \left( ( \vm{U}_{\text{TX}}^T \otimes \vm{U}_{\text{TX}}^H ) \vm{Z} ( \vm{U}_{\text{TX}}^{\ast} \otimes \vm{U}_{\text{TX}} ) \right)\IEEEnonumber\\
 &&\odot \left( \left( \vm{\Lambda}_{Q} \left( \vm{I}_{N_{\text{TX}}} + \rho \vm{\Lambda}_{\text{TX}} \vm{\Lambda}_{Q} \right)^{-1} \right)^T \otimes \left( \vm{\Lambda}_{Q} \left( \vm{I}_{N_{\text{TX}}} + \rho \vm{\Lambda}_{\text{TX}} \vm{\Lambda}_{Q} \right)^{-1} \right) \right) \big\}\IEEEnonumber\\
 &\stackrel{(\text{c})}{=}& \sum_{k=1}^{N_{\text{st}}} \sum_{l=1}^{N_{\text{st}}} \frac{ \left[ ( \vm{U}_{\text{TX}}^T \otimes \vm{U}_{\text{TX}}^H ) \vm{Z} ( \vm{U}_{\text{TX}}^{\ast} \otimes \vm{U}_{\text{TX}} ) \right]_{(k-1)N_{\text{TX}}+l, (k-1)N_{\text{TX}}+l}~ \lambda_{Q,k} \lambda_{Q,l} }{ \left( 1 + \rho \lambda_{\text{TX},k} \lambda_{Q,k} \right) \left( 1 + \rho \lambda_{\text{TX},l} \lambda_{Q,l} \right) }\IEEEnonumber\\
 &\stackrel{(\text{d})}{\leq}& \frac{1}{\rho^2} \sum_{k=1}^{N_{\text{st}}} \sum_{l=1}^{N_{\text{st}}} \frac{ \left[ ( \vm{U}_{\text{TX}}^T \otimes \vm{U}_{\text{TX}}^H ) \vm{Z} ( \vm{U}_{\text{TX}}^{\ast} \otimes \vm{U}_{\text{TX}} ) \right]_{(k-1)N_{\text{TX}}+l, (k-1)N_{\text{TX}}+l} }{ \lambda_{\text{TX},k} \lambda_{\text{TX},l} } .
 \eqnlab{MI_approx_LB_deriv}
\end{IEEEeqnarray}
In (a), we used \cite[Lemma~4.2.10]{Horn_Matrix_Analysis_2} as in \eqnref{app_second_order_approx_3}. In (b), we applied the identity $\tr{\vm{A} \vm{D}} = \tr{\vm{A} \odot \vm{D}}$ for matrices $\vm{A}$ and $\vm{D}$ of appropriate sizes, where $\vm{D}$ is diagonal. In (c), we made use of the fact that only the first $N_{\text{st}}$ elements on the diagonal of $\vm{\Lambda}_{Q}$ are non-zero. Finally, for (d), we note that $( \vm{U}_{\text{TX}}^T \otimes \vm{U}_{\text{TX}}^H ) \vm{Z} ( \vm{U}_{\text{TX}}^{\ast} \otimes \vm{U}_{\text{TX}} )$ is positive semidefinite.


\bibliographystyle{bib/IEEEtran}
\bibliography{bib/control,bib/IEEEabrv,bib/references}

\end{document}